\documentclass[aps,pra,twocolumn,superscriptaddress]{revtex4-2}
\usepackage{amsfonts}
\usepackage{bbding}
\usepackage{amssymb}
\usepackage[title]{appendix}
\usepackage{color}
\usepackage{amsmath}
\usepackage{graphicx}
\usepackage{subfigure}
\usepackage{txfonts}
\usepackage{bm}
\usepackage{ulem}
\usepackage{natbib}
\usepackage{calrsfs}
\DeclareMathAlphabet{\pazocal}{OMS}{zplm}{m}{n}
\newcommand{\Ss}{\pazocal{S}}

\newcommand{\Ns}{\pazocal{N}}
\newcommand{\Ls}{\pazocal{L}}
\newcommand{\Hs}{\pazocal{H}}
\newcommand{\Ps}{\pazocal{P}}
\newcommand{\Es}{\pazocal{E}}

\newcommand{\appsection}[1]{\section{\MakeUppercase{#1}}}

\usepackage[colorlinks,linkcolor=blue,citecolor=blue,urlcolor=blue]{hyperref}
\usepackage{orcidlink}

\preprint{APS/123-QED}

\begin{document}

\title{Quantum synchronization in an all-optical stroboscopic quantum simulator}

\author{Yan Li\orcidlink{0009-0007-7218-133X}}
\affiliation{Department of Physics, The Chinese University of Hong Kong, Shatin, New Territories, Hong Kong, China}

\author{Xingli Li\orcidlink{0000-0003-2339-6557}}
\email{xinglili@cuhk.edu.hk}
\affiliation{Department of Physics, The Chinese University of Hong Kong, Shatin, New Territories, Hong Kong, China}

\author{Wenlin Li\orcidlink{0000-0003-2032-0530}}
\email{liwenlin@mail.neu.edu.cn}
\affiliation{College of Sciences, Northeastern University, Shenyang 110819, China}

\begin{abstract}
In this work, we propose an all-optical stroboscopic scheme to simulate an open quantum system. By incorporating the tritter, consisting of a group of beam splitters, we find the emergence of spontaneous anti-phase synchronization in the steady state. To better understand the synchronization and entanglement properties within the system, we utilize the relative error measure and find the distribution of logarithmic negativity in parameter space shows similar structures with the results of synchronization measure. Finally, we derive the adjoint master equation corresponding to the system when the synchronization condition is satisfied and explain the existence of oscillations. In addition, we explore the effect of non-Markovianity on synchronization, and we find that it only slows down the time for the system to reach the steady state but does not change the synchronization properties of the steady state. Our work provides a promising scheme for experimental studies focused on synchronization and other nonequilibrium steady states. 
\end{abstract}
\date{\today}

\maketitle

\section{Introduction}

Synchronization, a phenomenon originally documented by Huygens~\cite{Huygens}, is a ubiquitous occurrence observed across various disciplines. Its influence extends across classical physics,
engineering, biological systems, and even social systems and economy~\cite{ikovskyRK01,strogatz2018nonlinear,pikovsky2012synchronization,Cai1993Oct, Glass2001Mar, Rosin2013Mar,Arenas2008Dec}. Moreover, in the past years, the investigation of synchronization has already been extended to the quantum domain, which refers to quantum synchronization. As a counterpart to classical synchronization in the realm of quantum mechanics, quantum synchronization has gained considerable attention in recent times~\cite{Galve2017Jun}.

Similar to classical synchronizations, quantum systems can be categorized based on the mechanisms that induce their synchronized behaviors: spontaneous synchronization and forced synchronization. 
Spontaneous synchronization occurs due to internal coupling within an interacting system, which is a result of the inherent interactions between the components of the system~\cite{Lee2014Feb, Xu2014Oct, Roulet2018Aug, Schmolke2022Dec, Schmolke2024Jan}; On the other side, forced synchronization, also known as entrainment~\cite{Adler1946Jun, Walter2014Mar, Roulet2018Jul, Murtadho2023Jul}, refers to the synchronized behavior that is induced by an external drive and leading to synchronization between the system and driving force. However, in contrast to the classical system, the presence of quantum properties, i.e., quantum correlations, decoherence, and superposition, significantly impacts quantum synchronization and gives rise to more fantastic phenomena~\cite{Roulet2018Aug}. Besides, the competition between the dissipation and the coherent dynamic also can lead to a synchronized behavior.

In recent times, numerous open quantum systems have been utilized to explore quantum synchronization from the theoretical perspective, which includes atomic ensembles~\cite{Xu2014Oct, Xu2015Mar, Hush2015Jun, Nadolny2023Nov}, superconducting circuits~\cite{Nongthombam2023Jan}, van
der Pol (vdP) oscillators~\cite{Lee2014Feb, Lee2013Dec, Weiss2017Apr, Jessop2020Feb}, optomechanical systems (OMSs)~\cite{Heinrich2011Jul, Ludwig2013Aug, Mari2013Sep, Ying2014Nov, Zhang2015Oct, Li2016Jun, Bemani2017Aug}, etc.
Furthermore, experimental realizations of quantum synchronization have been achieved in various platforms recently, including cold atoms~\cite{Laskar2020Jul}, trapped ions~\cite{Zhang2023Sep}, superconducting circuits~\cite{Vinokur2008Apr, Tao2024Jun}, and nuclear-spin system~\cite{Krithika2022Jun}.

As we have mentioned, the study of quantum synchronization has predominantly been conducted within the framework of the theory of open quantum systems, since a quantum system inevitably engages in interactions with its surrounding environment, leading to the occurrence of dissipation, decoherence, and other associated effects~\cite{breuer2002theory}. These characteristics classify the system as an open quantum system and can make a significant impact on the behavior and dynamics of the system.

In recent decades, a multitude of methods have emerged as effective means to model and address the equation of motion (EOM) for open quantum systems. Notable examples include Lindblad master equation~\cite{Lindblad1976Jun, Gorini1976May,breuer2002theory}, quantum jump method (also known as the Monte Carlo wave function method)~\cite{Dalibard1992Feb, Dum1992Apr, Molmer1993Mar}, and stochastic Schr\"odinger equations~\cite{Wiseman1993Jan, Plenio1998Jan, carmichael2009open, Daley2014Mar}. However, from a methodological standpoint, addressing the dynamics of system-bath interactions at a microscopic level is still inherently challenging. To this end, a framework known as a quantum collision model (QCM) or repeated interaction scheme has been proposed. The QCM draws inspiration from Boltzmann's original {\it Stosszahlansatz}~\cite{BrownSHPSB2009}, which describes the system only interacts with a tiny fraction of the degree of freedom of the environment at any given interval of evolution time. In practice, the tiny fraction of the degree of freedom of the environment generally is represented by a spin particle (e.g., qubit or qudit). Then, the sequential interactions between the system and a series of ``environmental particles'' will ultimately give rise to an open quantum system's dynamics. It has been proved that under certain conditions, QCM is equivalent to the Lindblad master equation. Moreover, ~\citet{Cattaneo2021Apr} proved that the QCM can efficiently simulate any multipartite Markovian quantum dynamics. Nowadays, the QCM has been widely employed in the study of various phenomena, including quantum thermodynamics~\cite{Strasberg2017Apr, Rodrigues2019Oct, Man2019Apr, Zhang2023Apr}, quantum non-Markovian dynamics~\cite{Ciccarello2013Apr, Kretschmer2016Jul, Camasca2021Feb}, quantum information~\cite{Beyer2018Mar, Heineken2021Nov, Korkmaz2023Jan}, and quantum synchronization~\cite{Karpat2019Jul, Karpat2021Jun, Li2023Apr, Mahlow2024May}. A comprehensive review of the current state-of-the-art in the QCM can be found in Ref.~\cite{Ciccarello2022Apr}.

Drawing upon the fundamental principles of the QCM, an all-optical stroboscopic quantum simulator has been proposed and implemented to investigate the non-Markovian dynamics by \citet{Jin2015Jan}. 
The system under investigation, along with the discrete degrees of freedom of the environment, can be effectively described in terms of frequency-degenerate quantum optical modes in this scheme. The interaction between those quantum optical modes is realized by a series of properly arranged passive linear-optical elements: beam splitters (BSs). 

The all-optical stroboscopic quantum simulator is a promising platform for the study of open quantum dynamics due to the advantages of linear optics experiments: integrability, high stability, low loss, and high controllability of device parameters. It is currently used in theoretical research to investigate non-Markovian dynamics~\cite{Jin2015Jan, Jin2018May, Li2024May}, and quantum information scrambling~\cite{Li2020Apr, Li2022Feb}.

Based on the notion of the QCM, we propose an all-optical stroboscopic quantum simulator to simulate an open quantum system within spontaneous quantum synchronization. After incorporating a multimode linear-optical element called a tritter into our QCM scheme, we are pleasantly surprised to discover that, under certain initial parameter settings, despite the system undergoing some dissipative dynamical processes, it will consistently exhibit time-dependent changes in the long-time limit, i.e., a persistent oscillation, rather than reaching a stable state. This behavior can be captured by the variation of Gaussian Wigner distribution at the different collision counts. To better explore and understand the oscillations and potential synchronization properties, we extend the well-known Mari's measure of synchronization~\cite{Mari2013Sep} with the relative error functions to satisfy our cases. Moreover, the dissipative coupling and resulting synchronization, triggered by the successive interactions of the system modes with the shared environmental modes, drive our investigation into the quantum correlations between the two system modes, particularly focusing on the entanglement. Consequently, we delve into the distribution of entanglement in the final state across parameter space by using the logarithmic negativity, and find that the system has higher synchronization properties along with subsystems with higher entanglement. In the final part of this work, we derive the corresponding adjoint master equations for the system to explain the existence of the oscillations and and discuss the effect of non-Markovianity on quantum synchronization. It is important to highlight that unlike previous research on quantum synchronization~\cite{Lee2014Feb}, our model operates within a linear framework, but we can still understand our work by using dark states. After deriving the Lindblad master equation with collective dissipation, we can find that the emergence of oscillations and synchronization in our system is attributed to the dynamic properties of collective dissipation, which identifies some specific states that serve as dark states of the dissipation operator while also acting as eigenstates of the Hamiltonian in the absence of detuning. This unique characteristic enables the system to exhibit a decoherence-free dynamic behavior, leading to the formation of synchronization.

This paper is structured as follows.  We start by introducing the fundamental information on our QCM scheme in Sec.~\ref{Sec:BackG}, including the basic introduction about the notion of the tritter and detailed setting of the all-optical QCM. The results are presented in Sec.~\ref{Sec:Res}, where we present the synchronization and entanglement properties of systems in parameter space. To gain a deeper understanding of this system, we reveal the corresponding adjoint master equation in Sec.~\ref{Sec:AME}. In addition, our research goes beyond the Markovian case to explore quantum synchronization in non-Markovian dynamics and present its results in Sec.~\ref{Sec:NMD}. We summarize in Sec.\ref{Sec:Summary}.

\section{Simulating open quantum systems by using an all-optical scheme}
\label{Sec:BackG}

 In this section, we aim to provide a detailed overview of the structure of our QCM scheme, as well as the necessary background information on the tritter and all-optical stroboscopic quantum simulator.

\subsection{The tritter}

Before delving into the specifics of the all-optical quantum simulator setup, we first briefly introduce a fundamental component known as the {\it tritter}~\cite{Zeilinger1993, Zeilinger1994, Greenberger1993Aug}, which plays an important role in our work.

\begin{figure}[!htpb]
    \centering
    \includegraphics[width=0.45\textwidth]{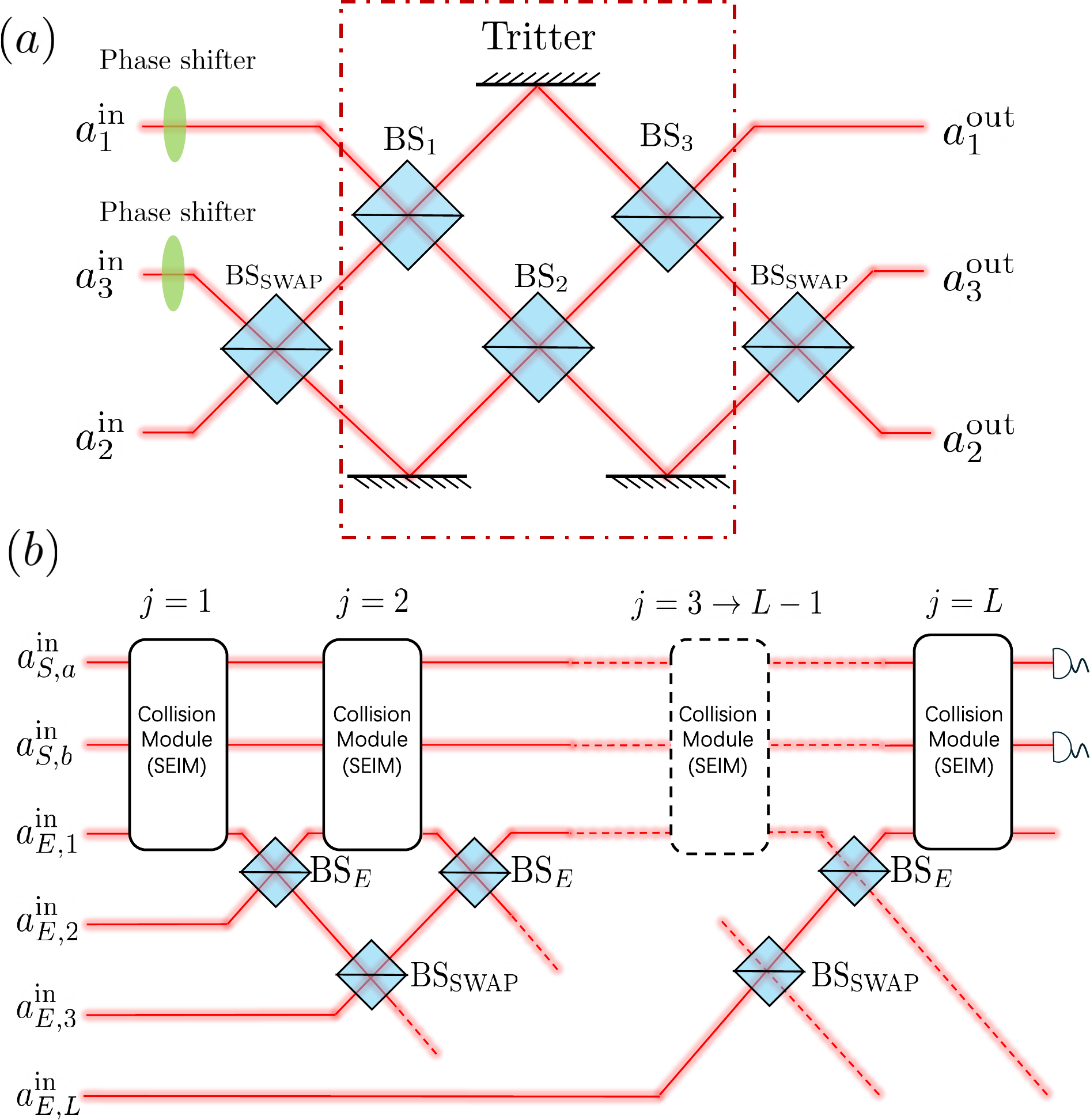}    
    \caption{\label{Fig:Sch} Schematic of the all-optical stroboscopic quantum simulator. (a) The system-environment interaction module (SEIM) describes the interaction between the system and environment optical modes, which is composed of a sequence of beam splitters, phase shifters, and mirrors. The box with the red dashed lines denotes a tritter, which contains three beam splitters. (b) The overall setup of the all-optical stroboscopic quantum simulator comprises two system optical modes, namely $a_{S,a}$ and $a_{S,b}$, along with a series of environment optical modes denoted as $a_{E,j=1\cdots L}$. The total number of collisions is represented by $L$. Each complete collision process consists of two consecutive parts. Firstly, there is the system-environment interaction, which is described by the SEIM. Secondly, there is the environment-environment interaction, which is captured by the beam splitter $\text{BS}_{\text{E}}$. The collision processes omitted in the middle are indicated by dashed lines, representing the continuation of the interaction sequence. Upon completing all collisions, the detectors are placed to measure and observe the properties of the system's optical modes. Additionally, the purpose of introducing the SWAP beam splitters in those above panels, which are denoted as $\text{BS}_{\text{SWAP}}=[0,1;1,0]$, is to exchange the positions of the two input modes. This strategic placement ensures that the optical paths do not intersect, thereby averting any potential misinterpretation or confusion in the schematic. However, it is important to note that the SWAP beam splitter is not essential in our specific numerical calculations.}
\end{figure}

The tritter is a kind of multimode linear-optical element that is generally defined as being able to
distribute all input optical modes into output ports with equal probability, i.e., symmetric multiport device~\cite{Greenberger1993Aug, Zukowski1997Apr, Campos2000Jun}. In this work, we preserve the arrangement of beam splitters to ensure the device structure of the tritter but extend beyond the symmetric multiport requirement, and this approach enabled us to achieve more intriguing and noteworthy results.

A multimode mixer can be effectively described by a unitary transformation known as the so-called scattering matrix, denoted as $\Ss$. It efficiently connects the input optical mode $a^{\text{in}}_n$ to output mode $a^{\text{out}}_n$ in a desired manner: $a^{\text{out}}_{m}=\sum_{n}\Ss_{mn}a^{\text{in}}_{n}$, where $a$ and $a^{\dagger}$ are annihilation and creation operators, respectively, follow the commutation relations $[a_{j},a^{\dagger}_{k}]=\delta_{jk}$. Experimentally, an arbitrary finite-dimensional unitary transformation can be implemented by utilizing combinations of passive linear-optical elements: beam splitters, phase shifters, polarizing beam splitter, etc~\cite{Reck1994Jul,Pan2012May}. The tritter, as its name suggests, operates on three optical modes simultaneously. However, the transformation matrix $\Ss_{\text{tritter}}$ for the tritter can be decomposed into a combination of specific beam splitters. As illustrated in the red dashed box of Fig.~\ref{Fig:Sch}(a), the decomposition is given by $\boldsymbol{a}^{\text{out}}=\Ss_{\text{tritter}}\boldsymbol{a}^{\text{in}}=\Ss^{(3)}_{\vartheta}\cdot\Ss^{(2)}_{\vartheta}\cdot\Ss^{(1)}_{\vartheta}\boldsymbol{a}^{\text{in}}$, with the first scattering matrix $\Ss^{(1)}_{\vartheta}$ characterizes the interferometer operating the $a_{1}$ and $a_{2}$ optical modes, 
\begin{equation}
\begin{pmatrix}
 a^{\text{out}}_{1}\\[0.25em]
 a^{\text{out}}_{2}\\[0.25em]
 a^{\text{out}}_{3}
\end{pmatrix}
=\Ss^{(1)}_{\vartheta}\begin{pmatrix}
 a^{\text{in}}_{1}\\[0.25em]
 a^{\text{in}}_{2}\\[0.25em]
 a^{\text{in}}_{3}
\end{pmatrix}=
\begin{pmatrix}
r_{1}&t_{1}&0\\[0.25em]
-t_{1}&r_{1}&0\\[0.25em]
0&0&1
\end{pmatrix}
\begin{pmatrix}
 a^{\text{in}}_{1}\\[0.25em]
 a^{\text{in}}_{2}\\[0.25em]
 a^{\text{in}}_{3}
\end{pmatrix},
\label{Eq:S1}
\end{equation}
and the second beam splitter is denoted by the scattering matrix $\Ss^{(2)}_{\vartheta}$, which characterizes the interferometer between the $a_{2}$ and $a_{3}$ optical modes,
\begin{equation}
\begin{pmatrix}
 a^{\text{out}}_{1}\\[0.25em]
 a^{\text{out}}_{2}\\[0.25em]
 a^{\text{out}}_{3}
\end{pmatrix}
=\Ss^{(2)}_{\vartheta}\begin{pmatrix}
 a^{\text{in}}_{1}\\[0.25em]
 a^{\text{in}}_{2}\\[0.25em]
 a^{\text{in}}_{3}
\end{pmatrix}
=
\begin{pmatrix}
1&0&0\\[0.25em]
0&r_{2}&t_{2}\\[0.25em]
0&-t_{2}&r_{2}\\
\end{pmatrix}
\begin{pmatrix}
 a^{\text{in}}_{1}\\[0.25em]
 a^{\text{in}}_{2}\\[0.25em]
 a^{\text{in}}_{3}
\end{pmatrix}.
\label{Eq:S2}
\end{equation}
The third beam splitter, like the first scattering matrix revealed in Eq.~(\ref{Eq:S1}), operates on the $a_{1}$ and $a_{2}$ optical modes. However, the parameters of the third one may differ
\begin{equation}
\begin{pmatrix}
 a^{\text{out}}_{1}\\[0.25em]
 a^{\text{out}}_{2}\\[0.25em]
 a^{\text{out}}_{3}
\end{pmatrix}
=\Ss^{(3)}_{\vartheta}\begin{pmatrix}
 a^{\text{in}}_{1}\\[0.25em]
 a^{\text{in}}_{2}\\[0.25em]
 a^{\text{in}}_{3}
\end{pmatrix}=
\begin{pmatrix}
r_{3}&t_{3}&0\\[0.25em]
-t_{3}&r_{3}&0\\[0.25em]
0&0&1
\end{pmatrix}
\begin{pmatrix}
 a^{\text{in}}_{1}\\[0.25em]
 a^{\text{in}}_{2}\\[0.25em]
 a^{\text{in}}_{3}
\end{pmatrix},
\label{Eq:S3}
\end{equation}
where the transformation matrices $\Ss^{(j)}_{\vartheta}$ are both $\vartheta$-dependent, directly controlled by the reflectivity $r_{j}=\cos{\vartheta_{j}}$ and transmissivity $t_{j}=\sin{\vartheta_{j}}$, which are satisfying $|r_{j}|^{2}+|t_{j}|^{2}=1$ and $\vartheta_{j}\in[0,\pi/2]$ to guarantee the preservation of SU(3) symmetry within the system~\cite{Zukowski1997Apr,Campos2000Jun}.

\subsection{The all-optical stroboscopic quantum simulator}

After establishing a foundation of the tritter's basic conception and functionality, we can now progress to introducing the overall setup of our specific research.

We utilize an all-optical stroboscopic quantum simulator to simulate an open quantum system, which consisting two system optical modes and a sequence of environment modes. As we previously mentioned, the stroboscopic interactions, also known as ``collisions", between the system and environment modes will result in a designed dissipative dynamic. According to the basic principles of the all-optical QCM, the interaction between two distinct optical modes is achieved through their mixture using a beam splitter. Therefore we design a so-called system-environment interaction module to describe this collision process.

As illustrated in Fig.~\ref{Fig:Sch}(a), the system-environment interaction module is built upon the foundation of a tritter and incorporates phase shifters~\cite{SWAP}. Through the mapping of each of the three modes into two system modes and one environment mode: $[a^{\text{in(out)}}_{1},a^{\text{in(out)}}_{3}, a^{\text{in(out)}}_{2}]\mapsto [a^{\text{in(out)}}_{S,a},a^{\text{in(out)}}_{S,b},a^{\text{in(out)}}_{E}]$, the scattering matrix of SEIM can be described as 
\begin{equation}
\Ss_{\text{SEIM}}=  
\begin{pmatrix}
r_3 & 0 & t_3 \\
0 & 1 & 0 \\
-t_3 & 0 & r_3 
\end{pmatrix}
\begin{pmatrix}
1 & 0 & 0 \\
0 & r_2  & t_2\\
0 & -t_2  & r_2 \\
\end{pmatrix}
\begin{pmatrix}
r_1 & 0 & t_1 \\
0 & 1 & 0 \\
-t_1 & 0 & r_1 
\end{pmatrix}
\begin{pmatrix}
e^{i \phi_a} & 0 & 0\\
0 & e^{i \phi_b} & 0\\
0 & 0 & 1
\end{pmatrix},
\end{equation}
where $\phi_{a(b)}$ is the phase shift angle for system mode $a_{S,a(b)}$.

To efficiently simulate an open quantum system, we need to introduce a sequence of system-environment mode interactions. In this way, we have to extend the SEIM and incorporate additional environment modes to build the complete simulator. The overall setup of the all-optical stroboscopic quantum simulator is demonstrated in Fig.~\ref{Fig:Sch}(b). It can be found that upon completing a system-environment mode interaction using SEIM, the ``new'' environment mode will be introduced through the environment-environment mode interaction beam splitter $\text{BS}_{E}$. This interaction results in the mixing of the ``old" environment mode with the ``new" environment mode. In addition, when there is no initial correlation between the system and the environment modes, no initial correlation and interaction between the environment modes themselves, and no repetitive interactions between the system and any of the environment modes, a so-called Markovian dynamics~\cite{Jin2015Jan, Jin2018May, Li2024May} will be obtained. Thus, when the parameters of the environment-environment mode interaction beam splitters $\text{BS}_{E}$ are set to full transmission, i.e., $r_{E}=0$, $t_{E}=1$, and the initial state of the joint system (system + environment) is a product state,
\begin{equation}
\varrho^{\text{in}}_{J} =  \varrho^{a}_{S}\otimes\varrho^{b}_{S}\otimes\prod_{j=1}^{L}\varrho_{E,j},
\label{Eq:Initial}
\end{equation}
where $L$ denotes the number of collision times (each collision introduces a new environment mode), the resulting dynamical process will exhibit Markovian characteristics. In addition to the Markovian dynamics, we also explore the non-Markovian scenario in this study. The corresponding results will be presented and discussed in the subsequent sections.

Upon completing the setup of the all-optical QCM, we can find that the dynamics in the time domain are discretized.
The dynamics can be represented as a sequence of discrete states undergoing multiple scattering matrices and can be captured in a stroboscopic manner. Consequently, we can express the continuous dynamical map equivalently in the following form,
\begin{equation}
\boldsymbol{a}^{\text{out}}(L)=\Ss_{J}(L)\cdot\boldsymbol{a}^{\text{in}}(L),  
\end{equation}
where the collection of the annihilation operators, denoted by $\boldsymbol{a}^{\text{in(out)}}(L)=[a_{S,a}^{\text{in(out)}},a_{S,b}^{\text{in(out)}},a_{E,1}^{\text{in(out)}},\cdots,a_{E,L}^{\text{in(out)}}]^{\top}$, stores the information of the system, and the superscript $\top$ represents the transpose operation.

The joint scattering matrix $\Ss_{J}(L)$ is a $L+2$-dimensional matrix that describes the process of mixing all optical modes, including both the SEIM and environment-environment interactions. Therefore, the specific form of the scattering matrix can be interpreted as follows~\cite{Mark}:
\begin{equation}
  \Ss_{J}(L) = \prod_{j=1}^{L}\Ss_{j},\quad \Ss_{j}= \Ss_{\vartheta}^{(3)}(j)\cdot\Ss_{\vartheta}^{(2)} (j)\cdot\Ss_{\vartheta}^{(1)}(j)\cdot\Ps,
\label{Eq:SMark}
\end{equation}
with each component can be written as 
\begin{equation}
\begin{aligned}
\Ss_{\vartheta}^{(1)}(j) 
&= 
\begin{pmatrix}
r_1 & 0 & t_1 & 0\\
0 & \mathbb{I}_j & 0 & 0\\
-t_1 & 0 & r_1 & 0\\
0 & 0 & 0 & \mathbb{I}_{L-j}
\end{pmatrix}, \\
\Ss_{\vartheta}^{(2)} (j) 
&= 
\begin{pmatrix}
1 & 0 & 0 & 0 & 0\\
0 & r_2 & 0 & t_2 & 0\\
0 & 0 & \mathbb{I}_{j-1} & 0 & 0\\
0 & -t_2 & 0 & r_2 & 0\\
0 & 0 & 0 & 0 & \mathbb{I}_{L-j}
\end{pmatrix},\\
\Ss_{\vartheta}^{(3)}(j) 
&= 
\begin{pmatrix}
r_3 & 0 & t_3 & 0\\
0 & \mathbb{I}_j & 0 & 0\\
-t_3 & 0 & r_3 & 0\\
0 & 0 & 0 & \mathbb{I}_{L-j}
\end{pmatrix}.
\end{aligned}
\end{equation}
The parameters $\vartheta_{1}$, $\vartheta_{2}$ and $\vartheta_{3}$, once set, will remain constants during the evolution. 

The form of the phase shifter matrix is fixed, 
\begin{equation}
\Ps = \begin{pmatrix}
e^{i \phi_a} & 0 & 0\\
0 & e^{i \phi_b} & 0\\
0 & 0 & \mathbb{I}_{L}
\end{pmatrix},
\end{equation}
where $\mathbb{I}_{j}$ is a $j\times j$-dimensional identity matrix, and the $\phi_{a(b)}$, as aforementioned, denote the angles of the phase shifter, which are likewise set and also held constants.

Since the Gaussianity of modes can be preserved in the linear optics setting, we restrict our discussions to the Gaussian domain. Considering a system consisting of $L$ bosonic modes, the Hilbert space of the total system $\Hs_{\text{tot}}$ can be represented by a tensor product form: $\Hs_{\text{tot}}=\otimes_{j=1}^{L}\Hs_{j}$. Here, each $\Hs_{j}$ denotes the infinite-dimensional Hilbert space of an optical mode. Then the quadrature operators for the $j$-th optical mode can be written as $q_{j}=(a^{\dagger}_{j}+a_{j})/\sqrt{2}$ and  $p_{j}=i(a^{\dagger}_{j}-a_{j})/\sqrt{2}$ and obey the canonical commutation relations, which is read as  
\begin{equation}
[q_{j},p_{k}]=i\Omega_{jk},\quad \Omega = \bigoplus_{j=1}^{N} \omega,\quad \omega = \begin{pmatrix}
 0 & 1\\
 -1 & 0
\end{pmatrix},    
\end{equation}
where $\Omega$ is a real canonical symplectic matrix. For any Gaussian state, all its information can be completely determined by a real symmetric matrix, namely, covariance matrix $\boldsymbol{\sigma}$. For a covariance matrix, which describes a {\it physical} Gaussian state if and only if the Robertson-Schr\"odinger uncertainty relation is satisfied~\cite{Simon1994Mar}. The relation is given by
\begin{equation}
  \boldsymbol{\sigma} + \frac{i}{2}\Omega \geq 0. 
\end{equation}
The covariance matrix for a system with $N$ bosonic modes can be completely characterized by its first and second statistical moments of the quadrature operator vector $\boldsymbol{R}$, 
\begin{equation}
 \boldsymbol{\sigma}_{jk} =  \langle\{R_{j}, R_{k}\}\rangle/2-\langle R_{j}\rangle\langle R_{k}\rangle, 
 \label{Eq:cov}
\end{equation}
where $\boldsymbol{R}=[q_{1},p_{1},\cdots, q_{L},p_{L}]^{\top}$, and $\{\cdot\}$ is the anticommutator, $\langle\cdot\rangle$ denotes the expectation value concerning the operator. It can be found that the covariance matrix for an $L$-mode Gaussian state is $2L$-dimensional~\cite{Weedbrook2012May}.

To investigate the dynamics within Gaussian channels, the characteristic function formalism is a useful tool for handling the elements of the covariance matrix~\cite{walls1994quantum}. In this formalism, the density operator of a system exhibits an equivalence with the characteristic function $\chi(\boldsymbol{\lambda})$ in presenting the probability distribution. Therefore, the multimode characteristic function of the input modes of the joint system can be expressed as 
\begin{equation}
   \chi^{\text{in}}_{J}(\boldsymbol{\lambda})= \text{tr}[D(\boldsymbol{\lambda})\varrho^{\text{in}}_{J}],
\end{equation}
where $D(\boldsymbol{\lambda})$ is the multimode Weyl displacement operator defined as $D(\boldsymbol{\lambda})=\otimes_{j}D(\lambda_{j})$, with $\boldsymbol{\lambda}=[\lambda_{S,a},\lambda_{S,b},\lambda_{E,1},\cdots,
\lambda_{E,L}]^{\top}$ is a complex vector and $D(\lambda_{j})=\exp{(\lambda_{j}a^{\dagger}-\lambda_{j}^{*}a_{j})}$ is the displacement operator for the $j$-th mode. However, to effectively capture the dynamics of the system, it is imperative to focus on the characteristic function of the output mode $\chi^{\text{out}}_{J}(\boldsymbol{\lambda})$, which is related to the inverse of the scattering matrix $\Ss_{J}^{-1}(L)=\Ss^{\dagger}_{J}(L)$ and input modes' characteristic function $\chi^{\text{in}}_{J}(\boldsymbol{\lambda})$ in the following way~\cite{Jin2018May},
\begin{equation}
\chi^{\text{out}}_{J}(\boldsymbol{\lambda})=\chi^{\text{in}}_{J} [\Ss^{-1}_{J}(L)\boldsymbol{\lambda}].
\label{Eq:outC}
\end{equation}
The characteristic function for the reduced density matrix associated with the desired mode can be obtained by performing a partial trace operation over all other bosonic modes. According to the theorem 2 shown in Ref.~\cite{Wang2007Aug}, the partial trace operation can be represented by substituting a specific complex vector, e.g., $\boldsymbol{\lambda}=[\lambda_{S, a},0,\cdots,0]^{\top}$ into Eq.~(\ref{Eq:outC}). In this manner, the reduced characteristic function of the output system mode $a_{S, a}^{\text{out}}$ will be expressed as 
\begin{equation}
 \chi^{\text{out}}_{S,a}(\lambda_{S,a})=   \chi^{\text{out}}_{J}(\boldsymbol{\lambda})|_{\boldsymbol{\lambda}=[\lambda_{S,a},0,\cdots,0]^{\top}}=\chi^{\text{in}}_{J}(\boldsymbol{M}\lambda_{S,a}),
 \label{Eq:ptcf}
\end{equation}
where $\boldsymbol{M}=[M_{1,1}, M_{1,2},\cdots, M_{1, L+1}, M_{1, L+2}]^{\top}$ denotes the elements of the first row of inversed scatting matrix $\Ss_{J}^{-1}(L)$~\cite{Jin2015Jan}. 

By utilizing the derivatives of the output mode's characteristic function at the origin of the complex plane, we can obtain symmetrically ordered moments of mode operators, i.e.,
\begin{equation}
\text{tr}\left[\rho^{\text{out}}_{J}[(a^{\dagger}_{j})^{n}a_{k}^{m}]_{\text{symm}}\right]=(-1)^{m}\frac{\partial^{n+m}}{\partial\lambda_{j}^{n}\partial(\lambda^{*}_{k})^{m}}\chi^{\text{out}}_{J}(\boldsymbol{\lambda})|_{\boldsymbol{\lambda}=0}.
\label{Eq:SOM}
\end{equation}
With the assistance of Eqs.~(\ref{Eq:cov}), (\ref{Eq:ptcf})-(\ref{Eq:SOM}), we can effectively construct the covariance matrix for the desired output pattern.

Following the overview of the fundamental setup of the all-optical QCM, we proceed to elaborate on the specific details of our initial configuration. In this particular work, we initialize each environmental optical mode as the coherent state, and the corresponding input characteristic function can be read as 
\begin{equation}
    \chi_{E,j}^{\text{in}}(\lambda_{E,j}) = \exp\left( -\frac{1}{2} |\lambda_{E,j}|^2  + \alpha_{E,j} \lambda_{E,j}^* + \alpha_{E,j}^* \lambda_{E,j}\right),
\end{equation}
where $\alpha_{E,j}$ is the complex displacement.  Without loss of generality, we can set the state of the environment to be identical for the convenience of calculations, i.e., $\chi^{\text{in}}_{E}(\lambda_{E})=\chi^{\text{in}}_{E,j}(\lambda_{E,j})$.

We consider the two system optical modes that do not have any correlation at the initial moment. Therefore, we initialize the two system modes as the separable general Gaussian state, e.g., $\varrho^{\beta}_{S}=D(\alpha_{\beta})S(\xi_{\beta},\varphi_{\beta})\varrho_{\text{th}}(n_{\beta})S^{\dagger}(\xi_{\beta},\varphi_{\beta})D^{\dagger}(\alpha_{\beta})$, where subscript $\beta=a,b$ for different system's optical modes. The operator $S(\xi_{\beta},\varphi_{\beta})$ denotes the squeezing operator with $\xi_\beta$ is the squeezing strength, $\varphi_\beta$ is the squeezing angle, operator $D(\alpha_{\beta})$ denotes the displacement operator with $\alpha_\beta$ is the complex displacement as aforementioned and $\varrho_{\text{th}}$ denotes the thermal state with $n_\beta$ being the thermal mean photon number.

In this way, the input characteristic function of the system optical mode $\beta$ will be given as~\cite{Marian2004Feb}
\begin{widetext}
\begin{equation}
\chi_{S,\,\beta}^{\text{in}}\left( \lambda_{S,\,\beta} \right)= \exp\left\{\left(n_{\beta}+\frac{1}{2}
\right)\left[ \frac{1}{2}\sinh{\xi_{\beta}}\left( e^{-i \varphi_{\beta}}\lambda_{S,\,\beta}^2+e^{i\varphi_{\beta}}\lambda_{S,\,\beta}^{*2}\right) -\cosh{\xi_\beta} |\lambda_{S,\,\beta}|^2 \right] -\alpha_{S,\,\beta} \lambda_{S,\,\beta}^{*} + \alpha_{S,\,\beta}^{*}\lambda_{S,\,\beta}\right\}.
\end{equation}
\end{widetext}
Following the definition of the initial joint input state given in Eq.~(\ref{Eq:Initial}), the joint input characteristic function will be
\begin{equation}
  \chi_{J}^{\text{in}}(\boldsymbol{\lambda})=\chi^{\text{in}}_{S,a}(\lambda_{S,a})\times\chi^{\text{in}}_{S,b}(\lambda_{S,b})\times\prod_{j=1}^{L}\chi^{\text{in}}_{E}(\lambda_{E}).
\end{equation}
With the help of Eqs.~(\ref{Eq:outC})-(\ref{Eq:SOM}), we can obtain the corresponding covariance matrix and stroboscopic dynamics of the system modes. The detailed expressions of the covariance matrices can be found in APPENDIX~\ref{APP:cov}.

In addition, we initialize the system modes in displaced squeezed states. While steady states in open quantum systems often do not depend on initial conditions, the entanglement dynamics in our system crucially depend on the initial state choice. Our analysis shows that within our framework of linear operations, coherent state initialization cannot generate entanglement between the system modes during the evolution. Displaced squeezed states, on the other hand, can generate entanglement via the beam splitter transformation~\cite{Braunstein2005Jun}. This choice of the displaced squeezed states enables us to investigate the dynamical evolution of entanglement, which is another key aspect of our study.

\section{Results}
\label{Sec:Res}

As a preliminary exploration, we initiate our discussions by showcasing the dynamic behavior of the system modes. In this pursuit, we consider the fact that any density matrix can be equivalently represented using a quasiprobability distribution, such as the Wigner function, defined over a real symplectic space, also known as the phase space~\cite{Weedbrook2012May}. The quasiprobability distribution can provide relatively more comprehensive information about the system than specific local expectations. With this in mind, we contemplate the advantages of utilizing the single-mode Gaussian Wigner distributions to capture the evolutionary processes of the two system modes separately.

The single-mode Gaussian Wigner function can establish a connection with the corresponding covariance matrix, allowing us to write it in the following manner~\cite{Wang2007Aug, walls1994quantum, Weedbrook2012May, Braunstein2005Jun}
\begin{equation}
  W(\boldsymbol{R})= \frac{1}{\pi\sqrt{\det{[\boldsymbol{\sigma}}]}} \exp{\left(-\frac{1}{2}(\boldsymbol{R}-\bar{\boldsymbol{R}})^{\top}\boldsymbol{\sigma}^{-1}(\boldsymbol{R}-\bar{\boldsymbol{R}})\right)}.
\end{equation}
where $\boldsymbol{R}=[q,p]$ represents the quadrature operator vector for a single-mode Gaussian state and $\bar{\boldsymbol{R}}$ denotes the displacement vector. 

\begin{figure}[!htpb]
    \centering
    \includegraphics[width=0.48\textwidth]{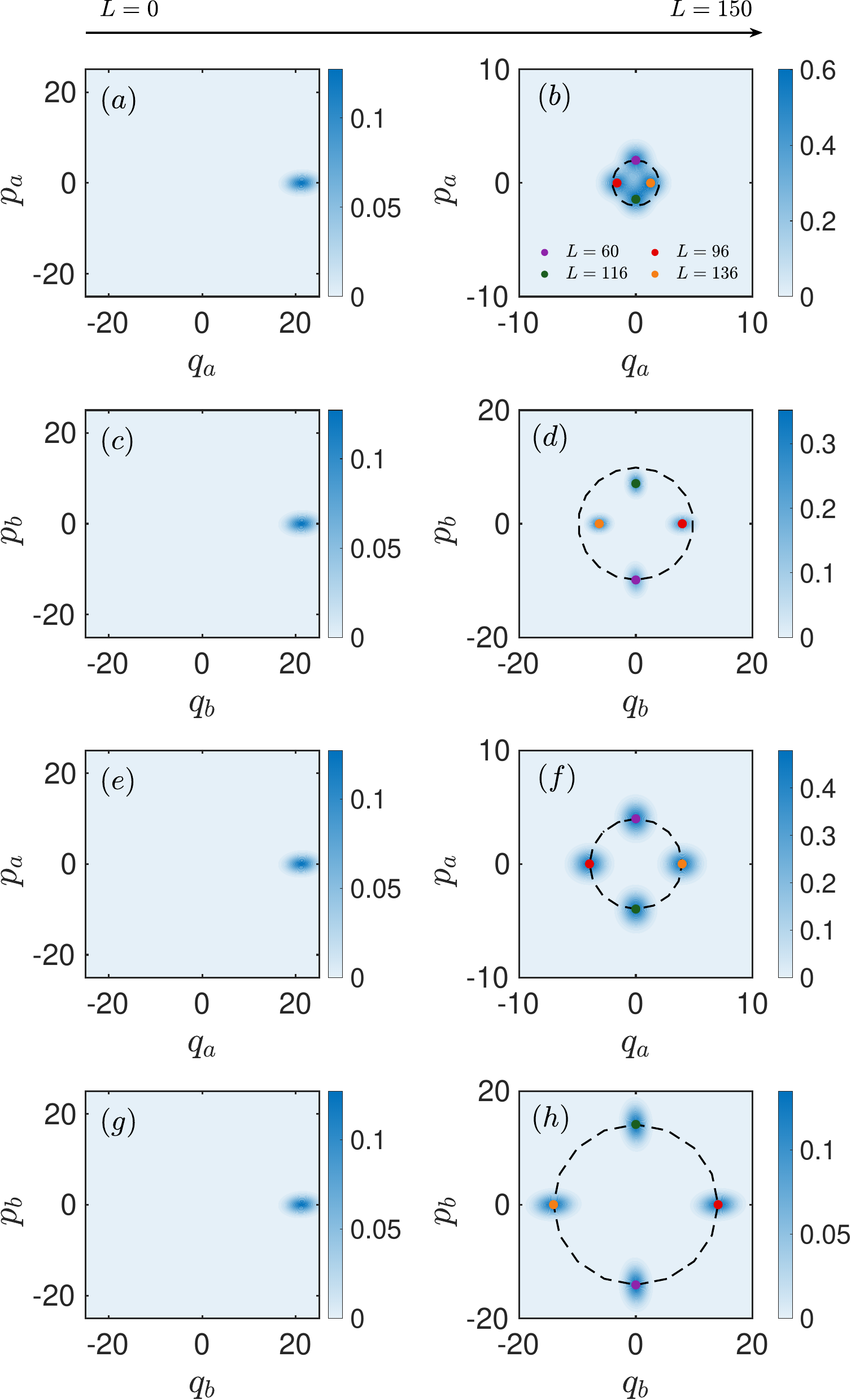}    
    \caption{\label{Fig:Wf} The phase space probability distribution of the two system's optical modes. (a)-(d) Unsynchronized case, where (a) and (c) are the Gaussian Wigner distributions at the initial moment, respectively. With the collisions progress, the Wigener distribution will exhibit variations in phase space. To illustrate those dynamic processes, we specifically showcase the Gaussian Wigner distribution for various numbers of collisions: $L = 60$ (purple dot), $L= 96$ (red dot), $L=116$ (green dot), and $L=136$ (orange dot). The dashed line indicates the displacement of the Wigner distribution from the phase space origin at $L = 60$, and the selected numbers remain consistent across (d), (f), and (h) panels. The parameters of beam splitters are $\{\vartheta_{1},\vartheta_{2},\vartheta_{3}\}/\pi=\{1/4,1/8,1/7\}$.   (e)-(h) Synchronized case, where (e) and (f) are also the distributions of the initial states. The parameter settings remain the same as the unsynchronized case, except for the configurations of the beam splitters, which are $\{\vartheta_{1},\vartheta_{2},\vartheta_{3}\}/\pi=\{1/4,1/8,1/4\}$. The remaining parameters are given as follows: $n_{\beta=a,b}=2$, $\xi_{\beta=a,b} =0.5$, $\varphi_{\beta=a,b}=0.1$, $\alpha_{\beta=a,b}=15$, $\phi_{\beta=a,b}=\pi/8$ and $\alpha_{E} = 0$. The total number of collisions is $L=150$. }    
\end{figure}

The results of the Gaussian Wigner distributions are shown in Fig.~\ref{Fig:Wf}.  In this exploration, we examine two distinct cases: the unsynchronized case,  represented by Fig.~\ref{Fig:Wf}(a)-(d), and the synchronized case,  represented by Fig.~\ref{Fig:Wf}(e)-(h), achieved by manipulating the angular values of the beam splitters. 

It is worth noting that in both cases, the two system optical modes are initially prepared in identical Gaussian states. To provide clarity on this initial condition ($L=0$), we present the initial Gaussian Wigner distributions in Fig.~\ref{Fig:Wf}(a) for the mode $a_{S,a}$ in the unsynchronized case, Fig.~\ref{Fig:Wf}(c) for the mode $a_{S,b}$ in the unsynchronized case, Fig.~\ref{Fig:Wf}(e) for the mode $a_{S,a}$ in the synchronized case, and Fig.~\ref{Fig:Wf}(g) for the mode $a_{S,b}$ in the synchronized case. Besides, the total number of collisions for both cases is $L=150$.

\begin{figure}[!htpb]
    \centering
    \includegraphics[width=0.48\textwidth]{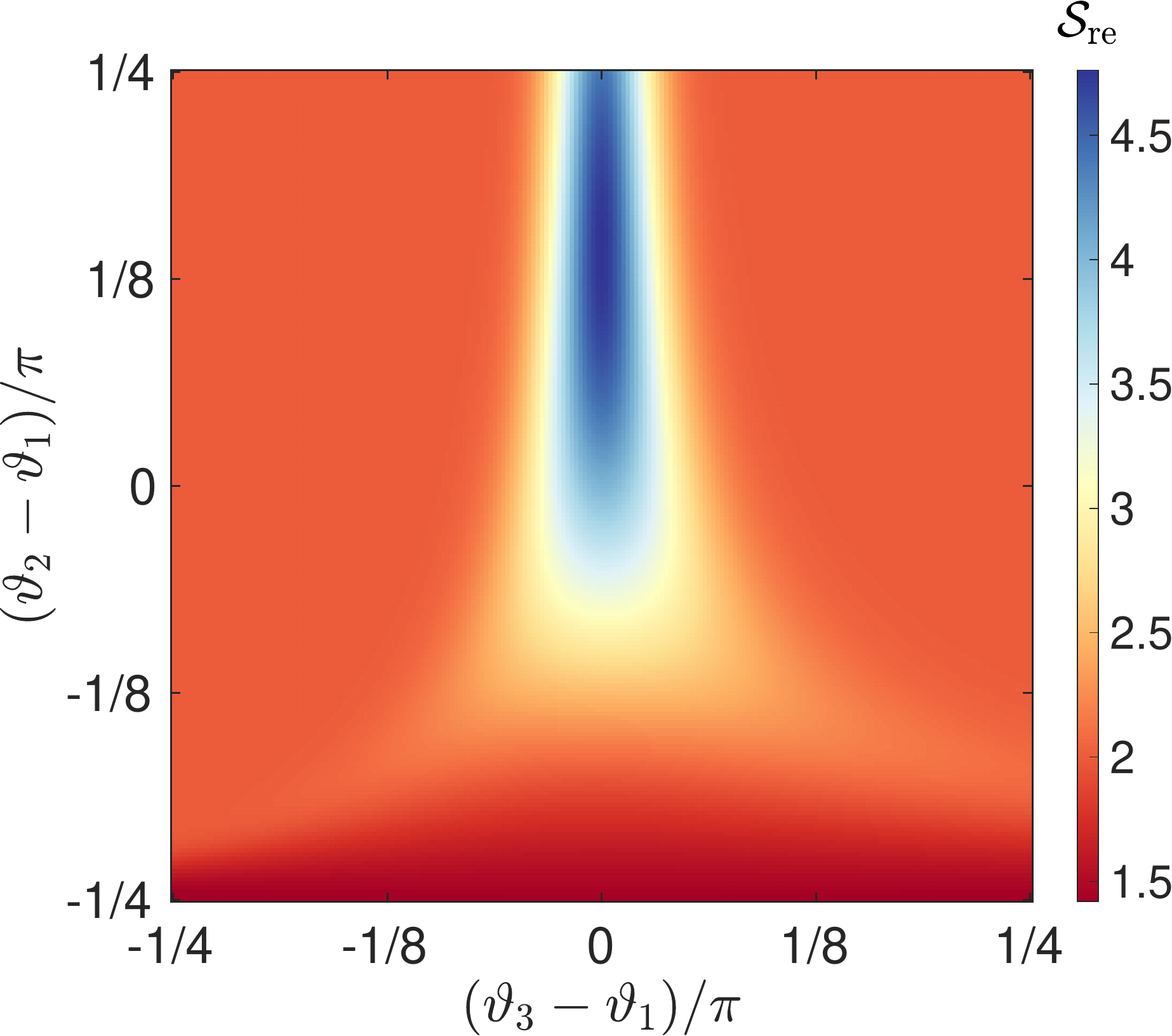}    
    \caption{\label{Fig:Sync} The result of the synchronization measure $\Ss_{\text{re}}$ variations in the $\vartheta_{3}-\vartheta_{2}$ parameter space with fixed $\vartheta_{1}=\pi/4$ to ensure symmetry in the parameter space. The other parameters for the initial system's optical modes are $n_{\beta=a,b}=0$, $\xi_{\beta=a,b}=2$, $\alpha_{\beta=a,b} = 1$, $\varphi_{\beta = a,b}= 2 $ and phase shift are $\phi_{\beta=a,b}=\pi/8$; for the environment optical mode are $\alpha_{E}=0$. The total number of collisions for each parameter point is $L=200$. }    
\end{figure}

Then we proceed to discuss the numerical results of the evolutionary process. In order to facilitate a comprehensive comparison between the unsynchronized and synchronized cases, we select four specific collision counts, e.g., $L=60$ (purple dot), $L=96$ (red dot), $L=116$ (green dot), $L=136$ (orange dot), and plot their corresponding Wigner distribution in a single phase space. Additionally, we compare the displacements at other collision counts with the displacements at $L=60$ to illustrate if there is some damping behavior in the system.

It can be easily found from Fig.~\ref{Fig:Wf}(b) and (d), which are the unsynchronized case, the absolute values of the displacements of the system modes $a_{S,a}$ and $a_{S,b}$ from the origin of the phase space gradually decrease with the collision number increase. This result suggests that, under the current parameters, the system will exhibit a damping oscillation and ultimately reach a steady state with zero displacement through the dissipative evolution process. In contrast, the system shows different behavior in Fig.~\ref{Fig:Wf}(f) and (h). We can find that the absolute values of the displacements of the system modes remain constants. The observation is indeed reinforced by the consistent positioning of the Gaussian Wigner distribution on the dashed circle, which has a fixed radius equivalent to the displacement of the collision count $L=60$. This consistency remains unchanged regardless of the number of collisions and indicates that the system is capable of sustaining a spontaneous and persistent oscillation within its dissipative dynamic process. In addition, by examining the locations of the two modes in the phase space at different collision counts, we are able to discern an anti-phase synchronization relationship between them.

\begin{figure*}[!htpb]
    \centering
    \includegraphics[width=1\textwidth]{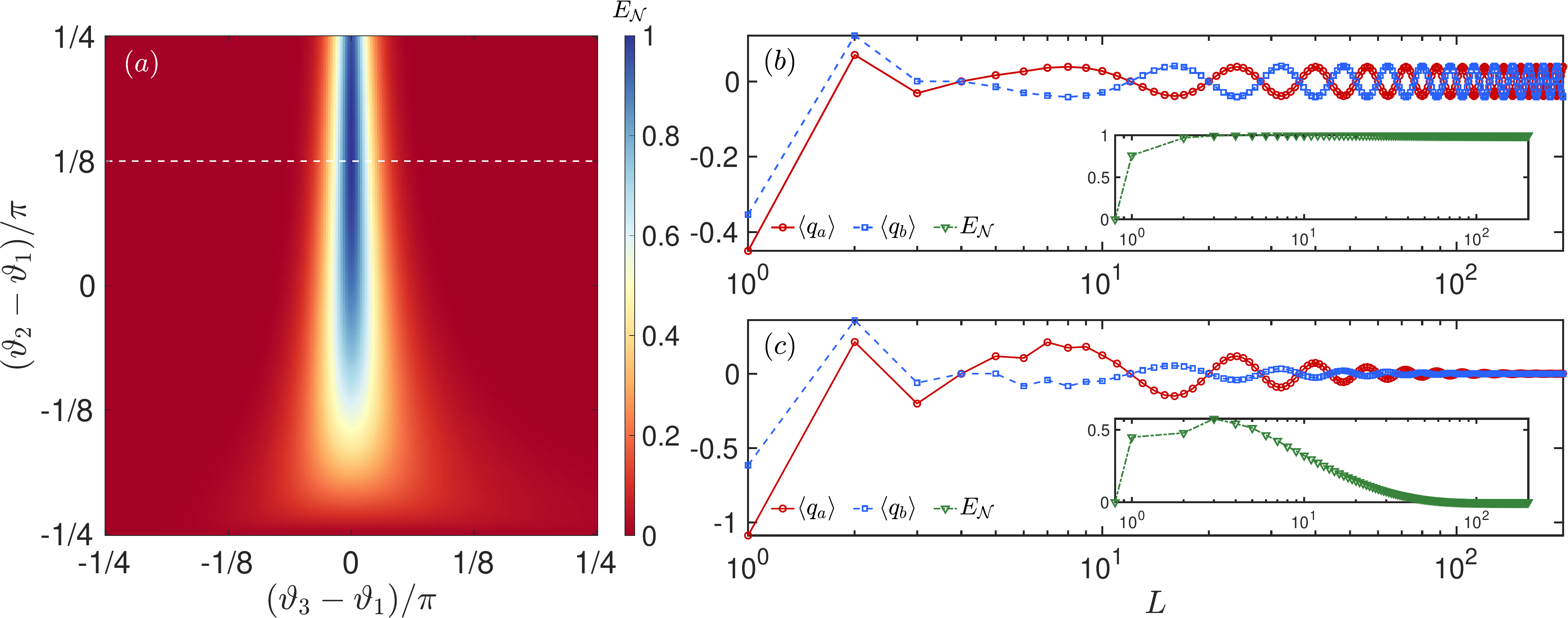}    
    \caption{\label{Fig:SM} (a) The logarithmic negativity $E_{\Ns}$ variations in the $\vartheta_{3}-\vartheta_{2}$ parameter space with fixed $\vartheta_{1}=\pi/4$. The total number of collisions is $L=200$. The other parameters for the system's optical modes are $n_{\beta=a,b}=0$, $\xi_{\beta=a,b}=2$, $\alpha_{\beta=a,b} = 1$, $\varphi_{\beta = a,b}= 2 $ and phase shift are $\phi_{\beta=a,b}=\pi/8$; for the environment optical mode are $\alpha_{E}=0$. The white dashed line corresponds to $\vartheta_{2}-\vartheta_{1}=\pi/8$, on which we have selected two specific parameter points for the specific discussion and shown in the following panels. (b)-(c) The expectation values of the position operators $\langle q_{\beta}\rangle$ as a function of the number of collision times $L$. The parameter setting for beam splitters are $(\vartheta_{2}-\vartheta_{1})/\pi=1/8$ and $(\vartheta_{3}-\vartheta_{1})/\pi=0$ for (b) and $(\vartheta_{2}-\vartheta_{1})/\pi=(\vartheta_{3}-\vartheta_{1})/\pi=1/8$ for (c). The insets for each panel show the logarithmic negativity varies with the collision times. The other parameters are the same with (a).}    
\end{figure*}

To gain deeper insights into the synchronization properties of the system's dynamics, we consider a specific measure of synchronization, which was first introduced by \citet{Mari2013Sep} in continuous variable (CV) systems, to quantify the degree of synchronization between different system modes.

The measure can be defined in the following form
\begin{equation}
  \Ss_{c}(L) \coloneqq  \langle q_{-}(L)^{2}+p_{-}(L)^2  \rangle^{-1},
\end{equation}
where $q_{-}(L)\coloneqq [q_{a}(L)+q_{b}(L)]/\sqrt{2}$ and $p_{-}(L)\coloneqq[p_{a}(L)+p_{b}(L)]/\sqrt{2}$ are the collision-dependent synchronization errors in our anti-phase synchronization case. Following the definition shown in Ref.~\cite{Mari2013Sep}, we can exclude the classical systematic error by shifting the variables $q_{-}$ and $p_{-}$ with their first moment, i.e.,
\begin{equation}
 q_{-}(L)\to q_{-}(L)-\langle q_{-}(L)\rangle,\quad p_{-}(L)\to p_{-}(L)-\langle p_{-}(L)\rangle.    
\end{equation}
where $\langle \cdot \rangle$ denotes the expectation value concerning the operator, as previously mentioned. The bounds for $\Ss_{c}$ lie within the range of [0,1], and $\Ss_{c}=1$ indicates the complete anti-phase synchronization.

In the long-time limit, i.e., $L\to\infty$, the system will eventually reach a steady state, and the steady value of $\Ss_{c}$ serves as the indicator of synchronization properties. However, it is important to note that in our all-optical QCM, the dimension of the joint system grows with the number of collisions. Consequently, we are limited to considering only a finite number of collisions for the evolution and analysis. As a result, when there is a weak damping process within the system under specific parameters, the final value of $\Ss_{c}$ with the given $L$ might tend to $\Ss_{c}=1$ (at this point the system does not actually reach steady state). This can make it challenging to accurately depict the damping case. To this end, we extend Mari's measure by introducing the notion of the relative error~\cite{Li2017Jul}, and reintroduce the following definition
\begin{equation}
  \Ss_{\text{re}}(L)  \coloneqq  \langle \Tilde{q}_{-}(L)^{2}+\Tilde{p}_{-}(L)^2  \rangle^{-1},  
\end{equation}
where the relative error functions are defined in the following forms $\Tilde{q}_{-}(L)\coloneqq [q_a(L)+q_b(L)]/\sqrt{2R_a(L)R_b(L)}$ and $\Tilde{p}_{-}(L)\coloneqq [p_a(L)+p_b(L)]/\sqrt{2R_a(L)R_b(L)}$ with $R_{\beta=a,b}(L)=\langle q_{\beta=a,b}(L)^2 + p_{\beta=a,b}(L)^2\rangle^{1/2}$. The relative measure lacks an upper bound, but provides a reasonably accurate description of the synchronization properties when the initial states are identical. 

The relative measure of synchronization is related to Mari's measure in such a way that $\Ss_{\text{re}}(L)=R_{a}(L)R_{b}(L)\Ss_{c}(L)$, and a comparison of the results obtained from the two measures is shown in APPENDIX \ref{APP:Com}.

We showcase the results of the measure of synchronization in the parameter space in Fig.~\ref{Fig:Sync}. Since the tritter has three free parameters, i.e., $\vartheta_{1}$,  $\vartheta_{2}$, and  $\vartheta_{3}$, we fix $\vartheta_{1}$ to reduce the parameter space and limit it into two dimensions. Moreover, to make the parameter space symmetric and to ensure the variation parameters cover all possible values, we let $\vartheta_{1}=\pi/4$. The results shown in Fig.~\ref{Fig:Sync} indicate that when the system satisfied the following conditions: (i) $|\vartheta_{3}-\vartheta_{1}|\to0$, (ii) $(\vartheta_{2}-\vartheta_{1})/\pi>-1/8$,  the dynamics will exhibit a high-degree of anti-phase synchronization. 

In our QCM scheme, the two system modes that do not directly interact with each other will bring about dissipative coupling by sequentially interacting with the environment modes, which may allow the system modes to generate quantum correlations, e.g., entanglement, in the dissipative evolution. Furthermore, we are interested in the relationship between quantum synchronization and entanglement, which prompts us to delve into studying entanglement variations within the system.

To quantify the entanglement between two subsystems, we introduce the notion of logarithmic negativity, denoted as $E_{\Ns}$. This measure serves as an extension of the so-called negativity to CV systems, it constitutes an upper bound to the distillable entanglement and is related to the entanglement cost under the positivity of the partially transposed (PPT) preserving operations~\cite{Vidal2002Feb}. The logarithmic negativity of a given quantum state $\varrho$ is defined as $E_{\Ns} = \ln||\Tilde{\varrho}||_{1}$, with $\Tilde{\varrho}$ is the partially transposed density matrix and $||\cdot||_{1}$ denotes the trace norm. Moreover, for any two-mode Gaussian state characterized by the covariance matrix $\boldsymbol{\sigma}$ as we mentioned previously, which can be specifically written as:
\begin{equation}
\boldsymbol{\sigma} =
\begin{pmatrix}
 A&C\\
 C^{\top}&B
\end{pmatrix}.    
\end{equation}
Then the logarithmic negativity can be reexpressed as the sum of all symplectic eigenvalues of the partially transposed matrix $\Tilde{\boldsymbol{\sigma}}=[A,-C;-C^{\top}, B]$ that are less than one. This simplifies the logarithmic negativity to $E_{\Ns} = \max[-\ln2\mu,0]$, where
\begin{equation}
 \mu = \sqrt{\frac{\Sigma-\sqrt{\Sigma^2-4\det[\boldsymbol{\sigma}]}}{2}},
\end{equation}
and $\Sigma=\det[A]+\det[B] - 2\det[C]$~\cite{Adesso2004Aug, Adesso2005Sep, Tserkis2017Dec}.

\begin{figure*}[!htpb]
    \centering
    \includegraphics[width=1\textwidth]{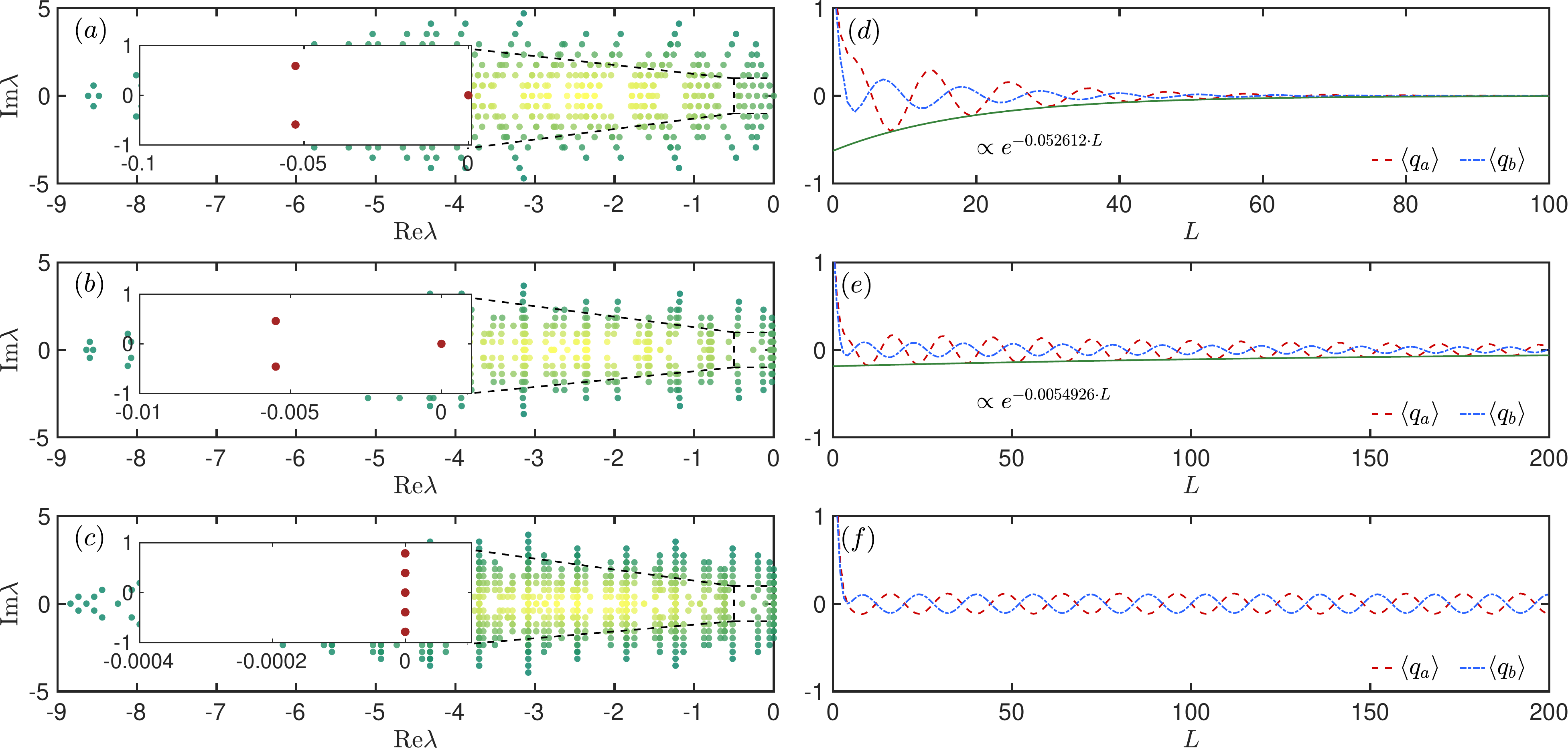}     
    \caption{\label{Fig:Decay} (a)-(c) The Liouvillian spectrum of the adjoint master equation (\ref{Eq:Liouvame}) with a truncated Hilbert dimension of $\dim[\Hs_{\text{trunc}}]=5$ in the complex plane. The insets are zoomed-in on the distribution of the complex eigenvalues near the original point $(\text{Re}\lambdaup,\text{Im}\lambdaup)=(0,0)$. (d)-(f) The results of the expectation value $\langle q_{\beta=a,b}\rangle$ with respect to the number of collisions, with the green solid line indicating the fitting curve obtained from the Liouvillian gap obtained in (a)-(c), respectively. The parameters for the initial system's optical modes are $n_{\beta=a,b}=0$, $\xi_{\beta=a,b}=2$, $\alpha_{\beta=a,b} = 1$, $\varphi_{\beta = a,b}= 2 $, and the phase shift are $\phi_{a}=\pi/8$ and $\phi_{b}=\pi/4$ for (a) and (d); $\phi_{a}=\pi/8$ and $\phi_{b}=\pi/6$ for (b) and (e); $\phi_{a}=\pi/8$ and $\phi_{b}=\pi/8$ for (c) and (f); The parameters for the environment optical mode are $\alpha_{E}=0$. As for the beam splitters are set as $\vartheta_{1}=\pi/8$, $\vartheta_{2}=\pi/4$ and $\vartheta_{3}=\pi/8$. The total number of collisions is $L=200$ for (d)-(f). }    
\end{figure*}

The results of logarithmic negativity are shown in Fig.~\ref{Fig:SM}(a). Upon comparing the distribution of results of entanglement in parameter space with the results depicted in Fig.~\ref{Fig:Sync}, a correlation between the high degree of entanglement and the relatively high level of synchronization within the parameter space can be observed. To better demonstrate this relationship, we will focus our discussion on the $(\vartheta_{2}-\vartheta_{1})/\pi=1/8$, which is indicated by the white dashed line in Fig.~\ref{Fig:SM}(a). Specifically, we select two parameter points along the line for detailed examination and analysis. In Fig.~\ref{Fig:SM}(b), the parameter setting is $(\vartheta_{3}-\vartheta_{1})/\pi=0$. Notably, we observe that the expectation values $\langle q_{\beta}\rangle$ of the position operators for both system modes gradually exhibit an anti-phase oscillatory pattern, with identical amplitudes, following an initial irregular evolution. Meanwhile, one can see from the inset of Fig.~\ref{Fig:SM}(b), initially, there is an absence of entanglement between the system modes. However, as the evolution progresses, we observe a gradual generation of entanglement, with it eventually reaching $E_{\Ns}=1$ when the dynamics achieve anti-phase synchronization. In contrast to the previous case, the dynamics of the system in Fig.~\ref{Fig:SM}(c) display a damping behavior when the parameter value is $(\vartheta_{3}-\vartheta_{1})/\pi=1/8$. This leads the system to eventually reach a trivial steady state. Simultaneously, the entanglement evolves from a state of non-existence to gradual generation. However, due to the system undergoing a trivial dissipation process, the entanglement is ultimately completely vanished. Those results further highlight a relationship between quantum synchronization and the generation of entanglement \cite{Roulet2018Aug}.

\section{Adjoint master equation}
\label{Sec:AME}

After conducting extensive numerical simulations that confirm the presence of oscillatory behaviors and anti-phase synchronization in our all-optical QCM, it is crucial to address a fundamental question: why does spontaneous synchronization occur in this system? In those previous works about QCM, it has been proved and repeatedly confirmed that, in the discrete systems, a Lindblad master equation can be derived from the QCM when the system undergoes a Markovian process and satisfies some specific initial conditions~\cite{Ciccarello2022Apr}. However, there is a lack of work on deriving the corresponding master equation from the all-optical QCM. Hence, our objective is to derive the appropriate master equation from the all-optical QCM and utilize it to explain the origin of oscillation.

The optical modes mixing process, as described by the scattering matrix, can be interpreted as an imaginary time evolution of the optical modes within an effective Hamiltonian $H$ in the Heisenberg picture, e.g., this can be demonstrated with the well-known beam splitter type Hamiltonian. Thus, we can assume the effective Hamiltonian and optical mode operators are all time-independent and start our discussion with the Heisenberg equation. Based on this assumption, the imaginary time-evolution unitary operator can be defined as $U=e^{-i\vartheta H}$, where the angle parameter $\vartheta$ denotes the imaginary time. Then we take a SEIM as an example, we can obtain the following mapping,
\begin{equation}
\begin{pmatrix}
 a^{\text{out}}_{S,a}\\[0.25em]
 a^{\text{out}}_{S,b}\\[0.25em]
 a^{\text{out}}_{E}    
\end{pmatrix}= 
\Ss_{\vartheta}^{(3)}\Ss_{\vartheta}^{(2)}\Ss_{\vartheta}^{(1)}\Ps\cdot\begin{pmatrix}
 a^{\text{in}}_{S,a}\\[0.25em]
 a^{\text{in}}_{S,b}\\[0.25em]
 a^{\text{in}}_{E}
\end{pmatrix} \mapsto
\left\{ \begin{array}{ll}
a^{\text{out}}_{S,a} = U^{\dagger} a^{\text{in}}_{S,a} U,\\[0.25em]
a^{\text{out}}_{S,b} = U^{\dagger} a^{\text{in}}_{S,b} U,\\[0.25em]
a^{\text{out}}_{E} = U^{\dagger} a^{\text{in}}_{E} U.
\end{array} \right.
\end{equation}
where $U=U_{\phi}U_{\text{tri.}}=U_{\phi}U_{\vartheta_{1}}U_{\vartheta_{2}}U_{\vartheta_{3}}$ and each component corresponds to a scattering matrix. We then can deduce the effective Hamiltonian corresponding to each unitary operator. (i) $U_{\phi}$ denotes the phase shift term that introduces phases to the system modes, defined by $a^{\text{out}}_{S,\,\beta}=e^{i\varphi_{\beta}}a^{\text{in}}_{S,\,\beta}$. Thus, the effective Hamiltonian for $U_{\phi}$ is $H_{\phi} = \phi_{a}N_{S,\, a} + \phi_{b}N_{S,\,b}$ where $N_{S,\,\beta}=a_{S,\,\beta}^{\dagger}a_{S,\,\beta}$ is the particle number operator for each system mode.  Besides, in the state of synchronization within the system, when $\phi_{a}=\phi_{b}$, the phase shifters' values are directly related to the oscillation frequency: $f=\phi_{a}/2\pi$. This point can be proved after we obtain the adjoint master equation. (ii) As for the tritter, each scattering matrix describes a two-mode beam splitter interaction [see Eqs.~(\ref{Eq:S1})-(\ref{Eq:S3})]. The beam splitter Hamiltonian can be formulated as $H=ab^{\dagger}+ba^{\dagger}$. Therefore, the effective Hamiltonian for $U_{\vartheta_{1}}$,$U_{\vartheta_{2}}$ and $U_{\vartheta_{3}}$ are as follows: $H_{\vartheta_{1}}=\vartheta_{1}(a_{S,a}a_{E}^{\dagger} + a_{S,a}^{\dagger}a_{E})$,
$H_{\vartheta_{2}}=\vartheta_{2}(a_{S,b}a_{E}^{\dagger} + a_{S,b}^{\dagger}a_{E})$, and $H_{\vartheta_{3}}$ has the same form as $H_{\vartheta_{1}}$, but with a different angle.

With the help of Baker-Campbell-Hausdorff formula~\cite{BCHf}, and recalling the parameter conditions under which the system appears to be in a persistent oscillation, we can approximately derive an effective Hamiltonian to describe the tritter: 
\begin{equation}
H_{\text{tri.}}\approx\Tilde{\vartheta}_{a}(a_{S,a}a_{E}^{\dagger} + a_{S,a}^{\dagger}a_{E})+ \Tilde{\vartheta}_{b}(a_{S,b}a_{E}^{\dagger} + a_{S,b}^{\dagger}a_{E}),  
\end{equation}
where $\Tilde{\vartheta}_{a}=\vartheta_{1} + \vartheta_{3}$ and $\Tilde{\vartheta}_{b}=\vartheta_{2}$. 

For the variation of the Weyl displacement operator of system modes with respect to the number of collisions (time), we can derive the following adjoint master equation,
\begin{equation}
\begin{aligned}
\frac{d}{dt} D(\lambda_{S})=&i[H_{\phi},D(\lambda_{S})] + o^{\dagger}D(\lambda_{S})o-\{o^{\dagger}o,D(\lambda_{S})\}/2,\\
 =&\Ls^{\dagger}[D(\lambda_{S})], 
 \label{Eq:adjEq}
\end{aligned}
\end{equation}
where the jump operator is defined as $o=\Tilde{\vartheta}_{a}a_{S,a} +\Tilde{\vartheta}_{b}a_{S,b}$, and $\{\cdot\}$ denotes the anti-commutator. The detailed derivations can be found in APPENDIX~\ref{App:AME}. The collective dissipation will lead to dissipative coupling~\cite{Lee2014Feb}, which explains the entanglement of the system modes without direct interaction. Moreover, this kind of collective dissipation has recently been experimentally realized in the framework of the QCM. By utilizing the IBM quantum computer, \citet{Cattaneo2023Mar} realized the collective dissipation for two-qubit system, i.e., $o=\sigma^{-}_{1} + \sigma^{-}_{2}$ with $\sigma^{-}_{j}=(\sigma^{x}_{j}-i\sigma^{y}_{j})/2$ and simulated superradiance and subradiance.

By vectorizing the operators in Hilbert space and mapping them into the Fock-Liouville space, we can decouple the Weyl displacement operator from the jump operator. When the dimension of the Hilbert space is the $\text{dim}[\Hs_{\text{Hil.}}]=L$, the adjoint Liouvillian superoperator can be written in a matrix form with the dimension being $\text{dim}[\Hs_\text{Liou.}]=L^2$~\cite{Minganti2018Oct},
\begin{equation}
\bar{\Ls}^{\dagger} = i[(H_{\phi}\otimes\mathbb{I})-(\mathbb{I}\otimes H^{\top}_{\phi})]+o^{\dagger}\otimes o^{\top}-o^{\dagger}o\otimes\mathbb{I}/2-\mathbb{I}\otimes o^{\top}o^{*}/2,
\label{Eq:Liouvame}
\end{equation}
where $\top$ and $*$ denote transpose and conjugate operations, respectively.  

It should be noted that although the $\bar{\Ls}^{\dagger}$ is a non-Hermitian matrix, in general, it can be diagonalizable, e.g., $\bar{\Ls}^{\dagger}\psi_{j}=c_{j}\psi_{j}$. Moreover, the eigenvalues $c_{j}$ are complex numbers and can be sorted in descending order by the absolute value of their real parts: $0=\text{Re}[c_{0}]\geq \text{Re}[c_{1}]\geq\cdots\geq \text{Re}[c_{L^2-1}]$. Then the adjoint master equation for a given operator $A$ can be written in the following way: $A(t)=A_{\text{ss}}+\sum_{j\neq0}^{L^2-1}g_{j}e^{c_{j}t}\psi_{j}$, where $g_{j}$ is the overlap between the given operator $A$ and $\psi_{j}$~\cite{Minganti2018Oct}. The absolute value of the eigenvalue $\text{Re}[c_{1}]$ is the slowest relaxation rate, which is referred to as the Liouvillian gap $c_{\text{gap}}$ and is also known as the asymptotic decay rate.

The damping behaviors of the collision number-dependent expectation values $\langle q_{\beta=a,b}\rangle_{L}$ can be described by the exponential decay with the Liouvillian gap, i.e., $\langle q_{\beta}\rangle_{L}\propto e^{-c_{\text{gap}} L}$. Therefore, we can leverage the dynamic behavior of $\langle q_{\beta}\rangle_{L}$, obtained through QCM, to estimate the corresponding decay rate. Then by comparing this decay rate with the Liouvillian gap, which is calculated under the same parameter settings as the dynamic results, we can evaluate the validity of the adjoint master equation~(\ref{Eq:adjEq}).

We take three parameter points as examples and present the results. In the first two rows of Fig.~\ref{Fig:Decay}, we present the distribution of the eigenvalues of $\bar{\Ls}^{\dagger}$ in complex space ((a) and (b)), as well as the evolution processes of $\langle p_{\beta}\rangle$ under the same parameter settings ((d) and (e)). Moreover, we select some data to fit the asymptotic decay process by using $\langle p_{\beta}\rangle_{L}$, which are highlighted with green lines. After comparing the fitted decay rates with the Liouvillian gap, we observe a strong agreement between them. This agreement strongly suggests the validity and accuracy of our adjoint master equation. 

Expanding on the previous findings, we consider the parameter point where the system can exhibit a complete anti-phase synchronization. We present its eigenvalues of $\bar{\Ls}^{\dagger}$ in Fig.~\ref{Fig:Decay}(c) and dynamic process in Fig.~\ref{Fig:Decay}(f). The pure imaginary eigenvalues, i.e., $\text{Re}[c_{j}]=0$, $\text{Im}[c_{j}]\neq0$ indicate that the system is capable of persistent oscillation~\cite{Buca2022Mar,Solanki2022Feb}.

Upon confirming the alignment between the adjoint master equation and the dynamic outcomes as indicated by the QCM, we can delve deeper into elucidating the emergence of oscillation and anti-phase synchronization within the system from the vantage point of the state obtained from the Lindblad master equation. In the Fock state basis, we consider the existence of a wave function that can be written in the following form: $|\psi\rangle=\sum_{n,m}c_{n,m}|n\rangle\otimes|m\rangle$, with the Fock states $|n(m)\rangle$ correspond to system modes $a(b)$, respectively. As a result, we will obtain the following results: $a_{S,a}|\psi\rangle=\sum_{n,m}c_{n,m}\sqrt{n}|n-1\rangle\otimes|m\rangle$ and $a_{S,b}|\psi\rangle=\sum_{n^\prime,m^\prime}c_{n^\prime,m^\prime}\sqrt{m^\prime}|n^\prime\rangle\otimes|m^\prime-1\rangle$. Then when the following relationships are satisfied: i) $n-1=n^\prime$ and $m=m^\prime-1$; ii) $\Tilde{\vartheta}_{a}c_{n,m}\sqrt{n}+\Tilde{\vartheta}_{b}c_{n^\prime,m^\prime}\sqrt{m^\prime}=0$, we can find the wave function $|\psi\rangle$ is a dark state for the dissipation part of Lindblad master equation,
\begin{equation}
 \frac{d}{dt}\varrho=-i[H_{\phi},\varrho] + o\varrho o ^{\dagger} - o^{\dagger}o\varrho/2 - \varrho o^{\dagger}o/2.
 \label{Eq:LME}
\end{equation}
with $\varrho=|\psi\rangle\langle\psi|$ and $o=\Tilde{\vartheta}_{a}a_{S,a}+\Tilde{\vartheta}_{b}a_{S,b}$. Moreover, in the case of $\phi_{a}=\phi_{b}$, the wave function $|\psi\rangle$ will also be the eigenstate of the Hamiltonian $H_{\phi}$. At this point, the system is decoherence-free. Moreover, since the dark states are the eigenstate of the Hamiltonian, we will have $|\psi(t)\rangle=e^{-i\phi_{a}t}|\psi(0)\rangle$, with $\phi_{a}=\phi_{b}$ is the premise. It can be found that the values of phase shifters can determine the oscillation frequency as mentioned before. Conversely, in scenarios where $\phi_{a}\neq\phi_{b}$, the wave function $|\psi\rangle$ will not be the eigenstate of $H_{\phi}$, and the composition of the state $|\psi\rangle$ will be disrupted by the unitary evolution component, rendering it a non-dark state susceptible to dissipation.

In addition, we can also make connections with the collective dissipation model~\cite{Bellomo2017Apr, Cabot2021May,Cattaneo2021May}. Given that the Lindblad master equation~(\ref{Eq:LME}) contains only the collective loss term,  we can approximately study the system by focusing on the discussion pertaining to the two lowest Fock states: $|0\rangle_{\beta}$ and $|1\rangle_{\beta}$. Through this approach, we are able to transform the adjoint master equation into a dissipative spin model, leading to the following master equation:
\begin{equation}
\frac{d}{dt}\varrho= -i[H,\varrho]+ O\varrho O^{\dagger}- O^{\dagger}O\varrho/2-\varrho O^{\dagger}O/2,
\end{equation}
where $H=\sum_{\beta=a,b}\phi_{\beta}\sigma^{+}_{\beta}\sigma^{-}_{\beta}$,$O=\Tilde{\vartheta}_{S,a}\sigma^{-}_{S,a}+\Tilde{\vartheta}_{S,b}\sigma^{-}_{S,b}$ with the annihilation and creation operators are transformed into spin-filp operator: $\sigma^{-}_{\beta}=|0\rangle\langle 1|_{\beta}$ and $\sigma^{+}_{\beta}=|1\rangle\langle 0|_{\beta}$. In this collective dissipation spin model, we can also explore spontaneous quantum synchronization and its connection with super- and subradiance~\cite{Cattaneo2023Mar, Bellomo2017Apr}.

\section{Non-Markovian dynamics}
\label{Sec:NMD}
The remarkable flexibility offered by the all-optical QCM allows for the transition of the dynamics from a Markovian to a non-Markovian process. This transition can be effortlessly achieved by introducing interactions within the environment modes. Consequently, in the final section, we provide a concise exploration of the effects of non-Markovian dynamics on synchronization.

\begin{figure}[!htpb]
    \centering
    \includegraphics[width=0.49\textwidth]{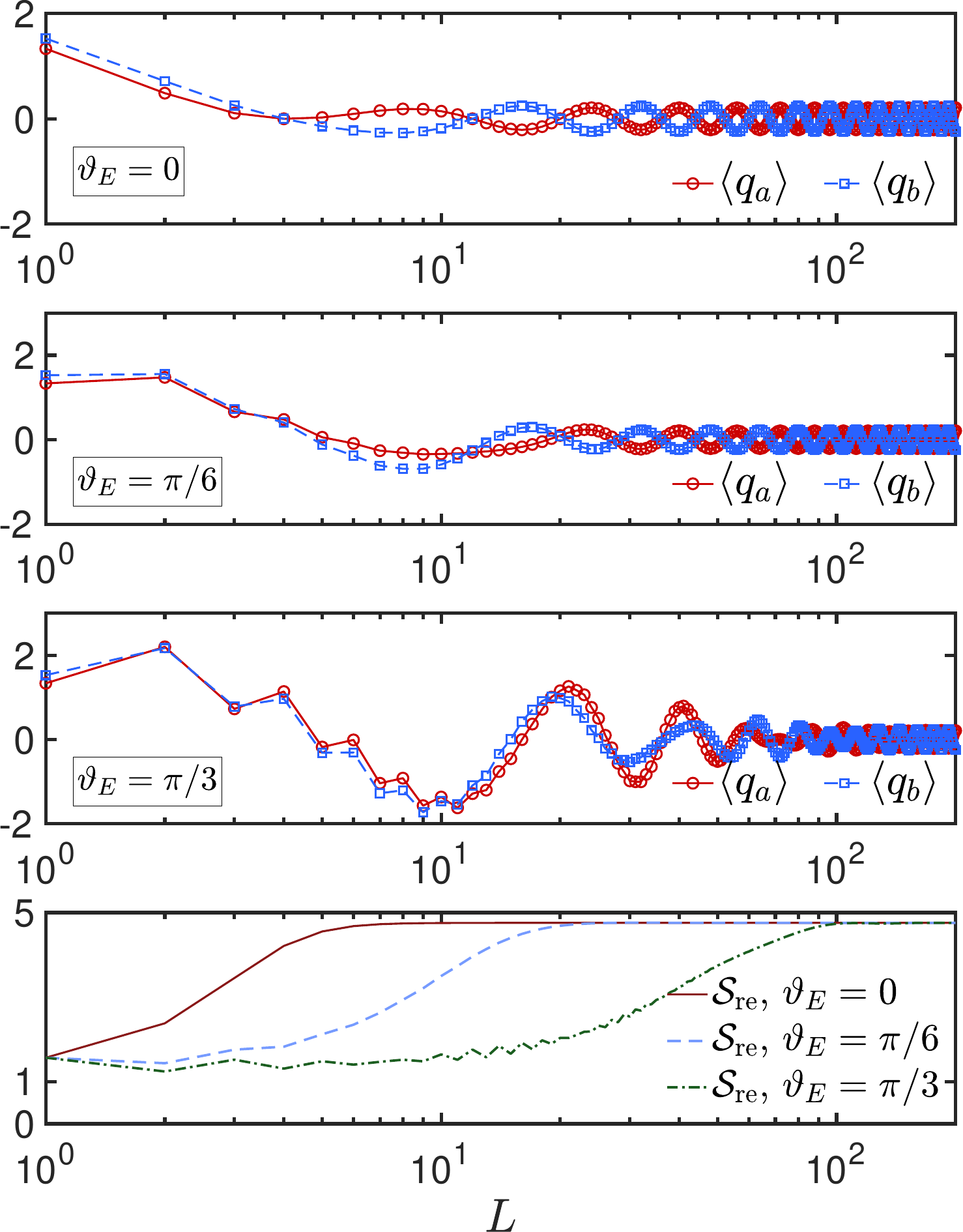}     
    \caption{\label{Fig:NM} The results of the expectation values $\langle q_{\beta=a,b}\rangle$ as a function of collision times $L$ are shown under different parameter settings of $\vartheta_{E}$, and the corresponding values of $\vartheta_{E}$ are indicated in the various panels. The bottom panel displays the results of the relative measure synchronization $\Ss_{\text{re}}$ over time for the above three cases. The parameter setting for tritters are $\{\vartheta_{1},\vartheta_{2},\vartheta_{3}\}/\pi=\{1/8,1/5,1/8\}$. The other parameters for the system's optical modes are $n_{\beta=a,b}=0$, $\xi_{\beta=a,b}=2$, $\alpha_{\beta=a,b} = 1$, $\varphi_{\beta = a,b}= 2 $ and phase shift are $\phi_{\beta=a,b}=\pi/8$; for the environment optical mode are $\alpha_{E}=0$. The total number of collisions is $L=200$. }  
\end{figure}

Recalling the scattering matrix shown in Eq.~(\ref{Eq:SMark}), we introduce the interactions between environment modes to construct $\Ss_{J}(L)$ for the non-Markovian case: $\Ss_{J}(L)=\prod_{j=2}^{L}[\Ss_{j}\cdot \Ss_{E}]\cdot \Ss_{1}$.
The scattering matrix $\Ss_{E}$ describes the internal interaction operation of the environment modes, which is represented by the beam splitter $\text{BS}_{E}$ in Fig.~\ref{Fig:Sch}(b), with a matrix expression of the form 
\begin{equation}
\Ss_{E} = \begin{pmatrix}
\mathbb{I}_{2} & 0 & 0 & 0 & 0\\
0 & \mathbb{I}_{j-2} & 0 & 0 & 0\\
0 & 0 & r_E & t_E & 0\\
0 & 0 & -t_E & r_E & 0\\
0 & 0 & 0 & 0 & \mathbb{I}_{L-j}
\end{pmatrix},
\end{equation}
where follows the previously established definition form: $r_{E}=\cos{\vartheta_{E}}$ and $t_{E}=\sin{\vartheta_{E}}$.

In Fig.~\ref{Fig:NM}, we present comparative results of the system's dynamics under various scenarios of environmental internal interactions, distinguished by differing parameter values of $\vartheta_{E}$. With the value of $\vartheta_{E}$ increasing, the system will undergo a transition from a Markovian dynamical process to a non-Markovian dynamical process~\cite{Jin2015Jan, Jin2018May, Li2024May}. Then we can find that this transition is accompanied by the emergence of more complex and intricate dynamical behavior during the initial stages of system evolution. However, it appears that the presence of non-Markovianity does not exert a substantial influence on the oscillation and synchronization properties of the system's steady state. To verify these observations, we provide the results of the relative measure of synchronization for those three parameter settings in the bottom panel of Fig.~\ref{Fig:NM}. The numerics provide further evidence that non-Markovianity influences the time it takes for the system to reach the steady state, but it does not change the ultimate steady-state properties.

\section{Summary}
\label{Sec:Summary}

In summary, our investigation focuses on spontaneous synchronization in an all-optical stroboscopic quantum simulator. Through the utilization of the theoretical framework based on the all-optical quantum collision model, we construct an open quantum system. After introducing the tritter into this model, our numerical results reveal the existence of persistent oscillatory behavior within the system modes, even in the absence of an external drive. Moreover, we extend Mari's measure of synchronization by incorporating relative error to gain a better understanding of the synchronization properties within this system. The distribution of the degree of synchronization in parameter space reminds us to explore the entanglement properties between the subsystems without correlations at the initial moment. To this end, we utilize the logarithmic negativity to characterize the final-state entanglement, and the result suggests a potential connection between entanglement and synchronization.

To gain a deeper understanding of the persistent oscillations observed within the system, we delve into the underlying principles using the mathematical perspective. After deriving the adjoint master equation corresponding to the system when it exhibits a higher degree of synchronization, we can obtain a concrete expression for the corresponding Liouvillian superoperator, whose eigenvalue spectrum provides us with crucial information to investigate the dynamics of the system. To ensure the validity of the accompanying master equation, we first compare the difference between the asymptotic decay rate obtained from fitting the curve of the operator expectation value over time and the Liouville gap. After confirming the validity of our derivation, we identify that under certain parameters the Liouvillian superoperator yields purely imaginary eigenvalues. This condition is a key requirement for the existence of persistent and stable oscillations in the system. Furthermore, we delve into the examination of the Lindblad master equation associated with the QCM to explain the oscillations within our linear system. Our analysis reveals the existence of some special states in the system, serving as both the dark states of the dissipation operator and eigenstates of the system's Hamiltonian. This unique characteristic allows these states to undergo purely coherent evolution within the master equation, leading to the emergence of oscillations. Additionally, we establish a connection between our model and the collective dissipative spin model with the consideration of the low-excitation subspace. Finally, we find that non-Markovianity has a specific effect on the system dynamics. It serves to delay the time required for the system to reach a steady state, but interestingly, it does not alter the final oscillatory properties of the system.

Despite extensive research on the spontaneous synchronization in various quantum systems and theoretical frameworks, including the quantum collision model~\cite{Karpat2019Jul, Karpat2021Jun, Li2023Apr, Mahlow2024May}, there remains a significant gap in the investigation of spontaneous synchronization in open quantum systems within an all-optical quantum framework. Our work aims to contribute to this research field, fill the existing void, and establish connections in other areas such as quantum time crystals. Moreover, our research provides a robust and versatile framework that can serve as a solid foundation for future experimental studies. This framework leverages the unique advantages of the all-optical setting, including high precision, controllability, and stability, which enhance the reliability and effectiveness of experimental investigations~\cite{Jin2015Jan, Jin2018May}.

\begin{acknowledgments}
This work has been supported by the National Natural Science Foundation of China under Grant No. 12304389 and by the Scientific Research Foundation of NEU under Grant No. 01270021920501*115.
\end{acknowledgments}

\appendix
\appsection{expression of covariance matrices}
\label{APP:cov}

In this appendix, we provide the detailed expression of the system's covariance matrix, denoted as $\boldsymbol{\sigma}$. The covariance matrix takes the form of a $4 \times 4$ matrix,
\begin{equation}
\boldsymbol{\sigma}=
\begin{pmatrix}
    A& C\\
    C^{\top}&B
\end{pmatrix}=
\begin{pmatrix}
A_{1,1} & A_{1,2} & C_{1,1} & C_{1,2} \\
A_{2,1} & A_{2,2} & C_{2,1} & C_{2,2} \\
C_{1,1} & C_{2,1} & B_{1,1} & B_{1,2} \\
C_{1,2} & C_{2,2} & B_{2,1} & B_{2,2}  
\end{pmatrix}.
\end{equation}
In our case, we have $A_{1,2} = A_{2,1}$, $B_{1,2} = B_{2,1}$, and $C_{1,2} = C_{2,1}$. It is worth noting that the values of the matrix elements in the covariance matrix are dependent on the collision number $L$. This dependency arises from the fact that the elements are interconnected with the elements of the inversed scattering matrix, i.e., $\Ss^{-1}(L)$, which itself is dependent on the number of collisions. However, in the interest of brevity and maintaining focus, we have chosen not to explicitly include the expression for the number of collisions in this covariance matrix. Moreover, to simplify the expression and improve clarity, we redefine the following variables: $n_{a}^{\text{th}}=n_{a}+\frac{1}{2}$, $n_{b}^{\text{th}}=n_{b}+\frac{1}{2}$, $\Tilde{M}_{1-2,1}=(1-|M_{1,1}|^2-|M_{2,1}|^2)/2$, $\Tilde{M}_{1-2,2}=(1-|M_{1,2}|^2-|M_{2,2}|^2)/2$, and $\Tilde{M}_{\text{sum}}= \frac{1}{2} \sum_{j=3}^{L+2} M_{j,1}^{*}M_{j,2}$. In this way, the specific patterns of the matrix elements can be given in the following form,
\begin{equation}
\begin{aligned}
A_{1,1}= & n_{a}^{\text{th}}\sinh \xi_a(\cos \varphi_a \operatorname{Re} [M_{1,1}^2]+\sin \varphi_a \operatorname{Im}[M_{1,1}^2])\\
&+n_{b}^{\text{th}}\sinh \xi_b(\cos \varphi_b \operatorname{Re}[M_{2,1}^2]+\sin \varphi_b \operatorname{Im}[M_{2,1}^2]) \\
& + n_{a}^{\text{th}} \cosh \xi_a |M_{1,1} |^2+n_{b}^{\text{th}} \cosh \xi_b|M_{2,1}|^2+\Tilde{M}_{1-2,1},\\
A_{2,2} = & -n_{a}^{\text{th}} \sinh \xi_a (\cos \varphi_a  \operatorname{Re} [M_{1,1}^2]+\sin \varphi_a \operatorname{Im}[M_{1,1}^2])\\
&-n_{b}^{\text{th}} \sinh \xi_b(\cos \varphi_b \operatorname{Re}[M_{2,1}^2]+\sin \varphi_b \operatorname{Im}[M_{2,1}^2])\\
& + n_{a}^{\text{th}} \cosh \xi_a |M_{1,1} |^2+n_{b}^{\text{th}} \cosh \xi_b|M_{2,1}|^2+\Tilde{M}_{1-2,1},\\
A_{1,2} = &n_{a}^{\text{th}} \sinh \xi_a (\sin \varphi_a \operatorname{Re}[M_{11}^2]-\cos \varphi_a \operatorname{Im}[M_{1,1}^2])\\
&+n_{b}^{\text{th}} \sinh \xi_b (\sin \varphi_b \operatorname{Re}[M_{2,1}^2]-\cos \varphi_b \operatorname{Im}[M_{2,1}^2]),\\
\end{aligned}    
\end{equation}
and
\begin{equation}
\begin{aligned}
B_{1,1} = & n_{a}^{\text{th}}\sinh\xi_a(\cos\varphi_a \operatorname{Re} [M_{1,2}^2]+\sin\varphi_a \operatorname{Im} [M_{1,2}^2])\\
&+n_{b}^{\text{th}}\sinh \xi_b(\cos \varphi_b \operatorname{Re}[M_{2,2}^2]+\sin \varphi_b \operatorname{Im}[M_{2,2}^2]) \\
& + n_{a}^{\text{th}}\cosh \xi_a |M_{1,2}|^2+n_{b}^{\text{th}}\cosh \xi_b|M_{2,2}|^2+\Tilde{M}_{1-2,2},\\
B_{2,2} = & -n_{a}^{\text{th}} \sinh \xi_a(\cos \varphi_a  \operatorname{Re} [M_{1,2}^2]+\sin \varphi_a \operatorname{Im} [M_{1,2}^2])\\
&-n_{b}^{\text{th}} \sinh \xi_b(\cos \varphi_b \operatorname{Re}[M_{2,2}^2]+\sin \varphi_b \operatorname{Im}[M_{2,2}^2]) \\
& + n_{a}^{\text{th}} \cosh \xi_a |M_{1,2} |^2+ n_{b}^{\text{th}} \cosh \xi_b|M_{2,2}|^2+\Tilde{M}_{1-2,2},\\ 
B_{1,2} = & n_{a}^{\text{th}}\sinh \xi_a (\sin \varphi_a \operatorname{Re}[M_{1,2}^2]-\cos \varphi_a \operatorname{Im}[M_{1,2}^2])\\
&+n_{b}^{\text{th}} \sinh\xi_b (\sin \varphi_b \operatorname{Re}[M_{2,2}^2]-\cos \varphi_b \operatorname{Im}[M_{2,2}^2]),\\
\end{aligned}   
\end{equation}
and
\begin{equation}
\begin{aligned}
C_{1,1} = & n_{a}^{\text{th}} \sinh \xi_a (\cos \varphi_a \operatorname{Re} [ M_{1,1}M_{1,2} ] + \sin \varphi_a \operatorname{Im} [ M_{1,1}M_{1,2}])\\
&+n_{b}^{\text{th}} \sinh \xi_b ( \cos \varphi_b \operatorname{Re} [ M_{2,1}M_{2,2} ] + \sin \varphi_b \operatorname{Im} [ M_{2,1}M_{2,2} ] )\\
&+n_{a}^{\text{th}} \cosh \varphi_a M_{1,1}^{*}M_{1,2} + n_{b}^{\text{th}} \cosh \varphi_b M_{2,1}^{*}M_{2,2} +  \Tilde{M}_{\text{sum}},\\
C_{2,2} = & -n_{a}^{\text{th}} \sinh \xi_a ( \cos \varphi_a \operatorname{Re} [ M_{1,1}M_{1,2} ] + \sin \varphi_a \operatorname{Im} [M_{1,1}M_{1,2}])\\
& - n_{a}^{\text{th}}\sinh \xi_b (\cos \varphi_b \operatorname{Re}[M_{2,1}M_{2,2} ] + \sin \varphi_b \operatorname{Im} [M_{2,1}M_{2,2}])\\
 & + n_{a}^{\text{th}} \cosh \varphi_a M_{1,1}^{*}M_{1,2} + n_{b}^{\text{th}}\cosh\varphi_b M_{2,1}^{*}M_{2,2} + \Tilde{M}_{\text{sum}},\\  
C_{1,2} = &  n_{a}^{\text{th}}\sinh \xi_a ( \sin \varphi_a \operatorname{Re} [ M_{1,1}M_{1,2}] - \cos \varphi_a \operatorname{Im} [M_{1,1}M_{1,2}])\\
&+n_{b}^{\text{th}} \sinh \xi_b ( \sin \varphi_a \operatorname{Re} [M_{2,1}M_{2,2}] - \cos \varphi_b \operatorname{Im} [M_{2,1}M_{2,2}]),\\
\end{aligned}
\end{equation}
where $M_{j,k}$ represents the element situated in the $j$-th row and $k$-th column of the inversed scattering matrix.

\appsection{Comparison of $\Ss_{c}(L)$ and $\Ss_{\text{re}}(L)$} 
\label{APP:Com}

\begin{figure}[!htpb]
    \centering
    \includegraphics[width=0.49\textwidth]{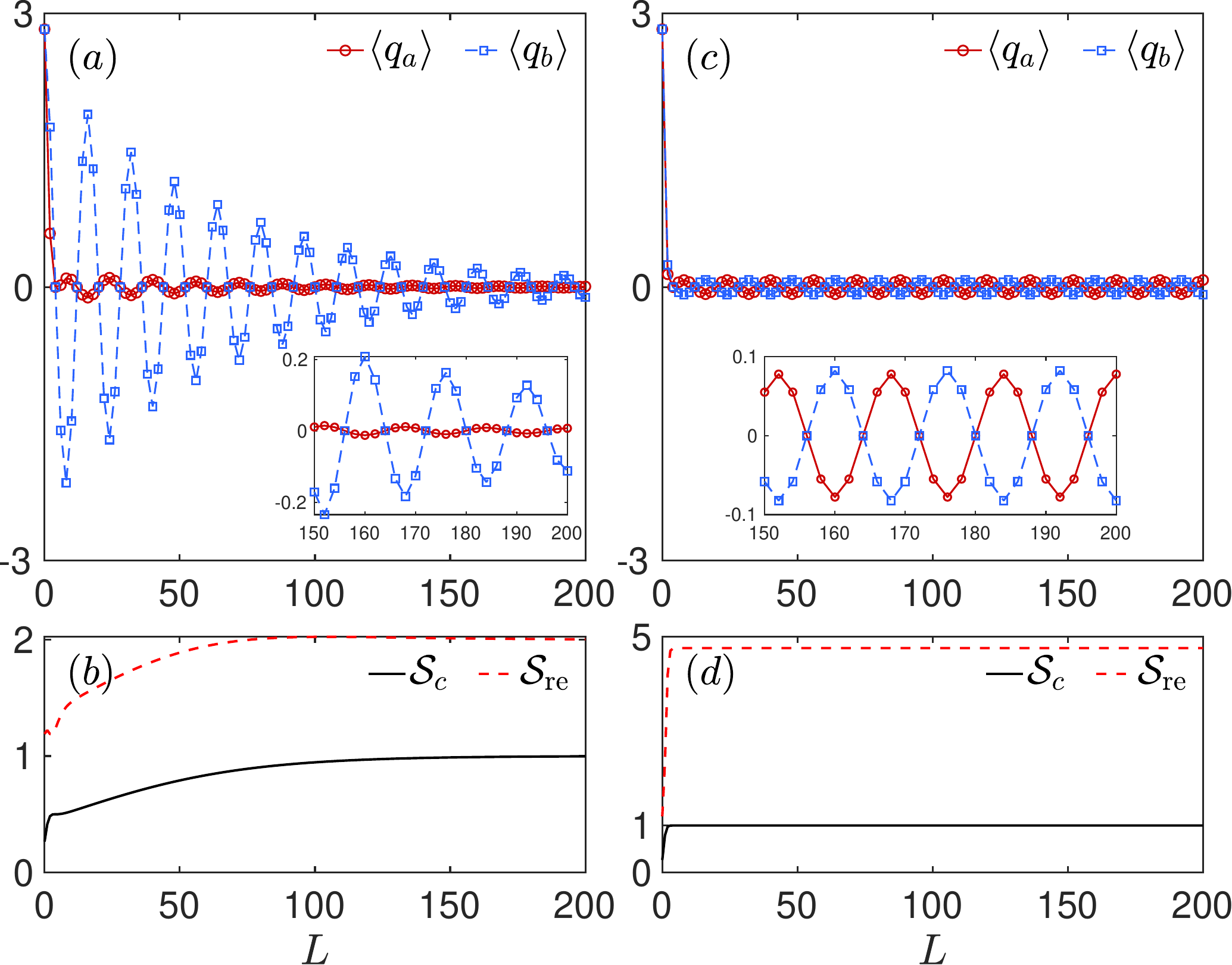}    
    \caption{\label{Fig:Appcom} Top panel: The expectation values of the position operators $\langle q_{\beta=a,b}\rangle$ as a function of collision times $L$: (a) for $(\vartheta_{2}-\vartheta_{1})/\pi=-1/6$, $(\vartheta_{3}-\vartheta_{1})/\pi=-1/5$ , and (c) for $(\vartheta_{2}-\vartheta_{1})/\pi=1/8$, $(\vartheta_{3}-\vartheta_{1})/\pi=0$.  Bottom panel: The measures of synchronization $\Ss_{c}$ and $\Ss_{\text{re}}$ vary with the collision times $L$: (b) for $(\vartheta_{2}-\vartheta_{1})/\pi=-1/6$, $(\vartheta_{3}-\vartheta_{1})/\pi=-1/5$ , and (d) for $(\vartheta_{2}-\vartheta_{1})/\pi=1/8$, $(\vartheta_{3}-\vartheta_{1})/\pi=0$. The other parameters used are identical to those in Fig.~\ref{Fig:Sync}. }    
\end{figure}

In the main text, we introduce the relative error functions to extend Mari's measure~\cite{Mari2013Sep} to make the measure more suitable for our situation.  This addresses the issue where Mari's measure tends to approach $\Ss_{c}\to1$ in the presence of damped oscillatory behavior, which may not accurately reflect the true level of quantum synchronization. In this way, we introduce the relative measure, incorporating the relative error functions, and present the comparison results in Fig.~\ref{Fig:Appcom}.

We show the expectation values of the position operators as a function of collision times $L$ in the top panels of Fig.~\ref{Fig:Appcom} with different parameter settings: (a) for damped oscillation and (c) for the perfect anti-phase synchronization. From Fig.~\ref{Fig:Appcom}(a), it is apparent that despite the consistent presence of oscillatory behavior in our limited simulations, the level of synchronization is expected to be lower in comparison to Fig.~\ref{Fig:Appcom}(c). However, the results shown in the bottom panels of Fig.~\ref{Fig:Appcom} indicate that Mari's measure has the same value in both cases, i.e., $\Ss_{c}\to 1$ hints the system is in a higher degree of anti-phase synchronization in both cases, which is contrary to what we observed in the top panel. On the other hand, the relative measure shows us a relatively more reasonable value.

\appsection{Derivation of the Adjoint Master Equation}
\label{App:AME}
We provide a detailed derivation of the adjoint master equation referenced in the main text in this part. Before proceeding with the specific details, we redefine the following notations to streamline the derivation: $D(\lambda_{S})=D(\lambda_{S,\,a})\otimes D(\lambda_{S,\,b})$ and $\varrho_{S}=\varrho^{a}_{S}\otimes\varrho^{b}_{S}$ and we fix the collision times is $L$. Recalling the multimode characteristic function of the input modes of the joint system $\chi^{\text{out}}_{J}$, we can have the following expressions~\cite{Jin2018May} 
\begin{equation}
\begin{aligned}
\chi^{\text{out}}_{S}(L)=&\chi^{\text{out}}_{J}(L)|_{\boldsymbol{\lambda}=[\lambda_{S,a},\lambda_{S,b},0,\cdots,0]},\\
=&\text{tr}[D(\boldsymbol{\lambda})U\varrho^{\text{in}}_{J}U^{\dagger}]|_{\boldsymbol{\lambda}=[\lambda_{S,a},\lambda_{S,b},0,\cdots,0]},\\
=&\text{tr}[U^{\dagger}D(\boldsymbol{\lambda})U\varrho^{\text{in}}_{J}]|_{\boldsymbol{\lambda}=[\lambda_{S,a},\lambda_{S,b},0,\cdots,0]},\\
=&\Es_{L}[\Es_{L-1}[\cdots\Es_{2}[\Es_{1}[ D(\lambda_{S})]]]],\\
=&\Es^{L}[D(\lambda_{S})],
\end{aligned}
\end{equation}
where $U=\prod_{j}^{L}U_{j}$ with $U_{j}=U_{\phi}U_{\text{tri.},j}$ are the unitary operators for the different processes. Following the standard procedure in the works related to QCM~\cite{Ciccarello2022Apr}, we define the quantum mapping $\Es_{j}[\cdot]$ in the above equation, which can be written as $\Es_{j}[D(\lambda_{S})]=\text{tr}[U_{j}^{\dagger}D(\lambda_{S})U_{j}\varrho_{S}\otimes \varrho_{E,j}]_{\boldsymbol{\lambda}=[\lambda_{S,a},\lambda_{S,b},0]}$. Then we can have the relationship between the change in the characteristic function and the collision (time),
\begin{equation}
\begin{aligned}
(\chi^{\text{out}}_{S})^{'}=&(\Es_{j}-\mathbb{I})[D(\lambda_{S})],\\
=&\text{tr}[\{U_{j}^{\dagger}D(\lambda_{S})U_{j}-D(\lambda_{S})\}\varrho_{S}\otimes \varrho_{E,j}].     
\end{aligned}
\end{equation}
After applying the Taylor expansion to those unitary evolution operators, we can approximate $U_{\text{tri.},j}$ as $\mathbb{I}-iH_{\text{tri.},j}+H_{\text{tri.},j}^2/2$ and $U_{\phi}$ as 
$\mathbb{I}-iH_{\phi}$. We then recall the initial conditions for the state of the environment in the numerical simulations: $\alpha_{E,j}=0$, we will have $\text{tr}[H_{\text{tri.},j}\,\varrho_{E,j}]=0$. Neglecting the higher order terms and the constant term, we arrive at 
\begin{equation}
\begin{aligned}
(\chi^{\text{out}}_{S})^{'}=&\text{tr}[\{-D(\lambda_{S})\frac{H^{2}_{\text{tri.},j}}{2}-iD(\lambda_{S})H_{\phi}+H_{\text{tri.},j}D(\lambda_{S})H_{\text{tri.},j}\\
&-\frac{H^{2}_{\text{tri.},j}}{2}D(\lambda_{S})+iH_{\phi}D(\lambda_{S})\}\varrho_{S}\otimes\varrho_{E,j}].\\
\end{aligned}    
\end{equation}
We expand the second-order term
\begin{equation}
\begin{aligned}
H_{\text{tri.},j}^{2} =&[(\Tilde{\vartheta}_{a}a_{S,a} + \Tilde{\vartheta}_{b}a_{S,b})a^{\dagger}_{E,j} + (\Tilde{\vartheta}_{a}a^{\dagger}_{S,a} + \Tilde{\vartheta}_{b}a^{\dagger}_{S,b})a_{E,j} ]^2 \\
=&[(\Tilde{\vartheta}_{a}a_{S,a} + \Tilde{\vartheta}_{b}a_{S,b})a^{\dagger}_{E,j}]\cdot[(\Tilde{\vartheta}_{a}a^{\dagger}_{S,a} + \Tilde{\vartheta}_{b}a^{\dagger}_{S,b})a_{E,j}] \\
&+ [(\Tilde{\vartheta}_{a}a^{\dagger}_{S,a} + \Tilde{\vartheta}_{b}a^{\dagger}_{S,a})a_{E,j}]\cdot [(\Tilde{\vartheta}_{a}a_{S,a} + \Tilde{\vartheta}_{b}a_{S,b})a^{\dagger}_{E,j}]\\
&+ [(\Tilde{\vartheta}_{a}a_{S,a} +\Tilde{\vartheta}_{b}a_{S,b})a^{\dagger}_{E,j}]^2 + [(\Tilde{\vartheta}_{a}a^{\dagger}_{S,a} + \Tilde{\vartheta}_{b}a^{\dagger}_{S,b})a_{E,j}]^2.\\
\end{aligned}
\end{equation} 
Still with the help of the initial conditions of the environment, we will have $\text{tr}[a_{E,j}^{2}\varrho_{E,j}]=\text{tr}[(a_{E,j}^{\dagger})^{2}\varrho_{E,j}]=\text{tr}[a_{E,j}^{\dagger}a_{E,j}\varrho_{E,j}]=0$, and $\text{tr}[a_{E,j}a_{E,j}^{\dagger}\varrho_{E,j}]=1$. We can define the jump operator $o$ as 
$o=\Tilde{\vartheta}_{a}a_{S,a} +\Tilde{\vartheta}_{b}a_{S,b}$, and have the following expression
\begin{equation}
\begin{aligned}
(\chi^{\text{out}}_{S})^{'}=&\text{tr}[\{iH_{\phi}D(\lambda_{S})-iD(\lambda_{S})H_{\phi}+o^{\dagger}D(\lambda_{S})o\\
&-D(\lambda_{S})\frac{o^{\dagger}o}{2}-\frac{o^{\dagger}o}{2}D(\lambda_{S})\}\varrho_{S}],\\
=&\text{tr}[\Ls^{\dagger}[D(\lambda_{S})]\varrho_{S}].\\
\end{aligned}    
\end{equation}
The above equation demonstrates that the change in the displacement operator, as a function of the number of collisions, can be expressed using the adjoint master equation.

\bibliography{Ref.bib}

\begin{thebibliography}{97}
\bibitem [{SWA()}]{SWAP}%
  \BibitemOpen
  \href@noop {} {}\bibinfo {howpublished} {The SWAP beam splitter setup prevents confusion caused by direct crossing of optical paths when interpreting the collision model. By rearranging the light modes within the tritter, specifically by exchanging the order of optical modes $a_{2}$ and $a_{3}$, the need for the SWAP beam splitter in the joint scattering matrix can be eliminated. Furthermore, it is not necessary to include the SWAP beam splitter in actual experiments.}\BibitemShut {Stop}%
\bibitem [{Mar()}]{Mark}%
  \BibitemOpen
  \href@noop {} {}\bibinfo {howpublished} {Since the environment modes do not interact with each other in the Markovian case, the scattering matrix $\Ss_{J}(L)$ can be simplified as shown in Eq.~\eqref{Eq:SMark}.}\BibitemShut {Stop}%
\end{thebibliography}

\end{document}